\documentclass[nojss]{jss}
\usepackage{amsmath}
\usepackage{amssymb}
\usepackage{bm}
\usepackage{tikz}
\usepackage{tikz-qtree}

\author{Justin Angevaare\\University of Guelph \And 
        Zeny Feng\\University of Guelph \And
        Rob Deardon\\University of Calgary}
\title{\pkg{Pathogen.jl}: Infectious Disease Transmission Network Modelling with \proglang{Julia}}

\Plainauthor{Justin Angevaare, Zeny Feng, Rob Deardon} 
\Plaintitle{Pathogen.jl: Infectious Disease Transmission Network Modelling with Julia} 

\Abstract{
We introduce \pkg{Pathogen.jl} for simulation and inference of transmission network individual level models (TN-ILMs) of infectious disease spread in continuous time. TN-ILMs can be used to jointly infer transmission networks, event times, and model parameters within a Bayesian framework via Markov chain Monte Carlo (MCMC). We detail our specific strategies for conducting MCMC for TN-ILMs, and our implementation of these strategies in the \proglang{Julia} package, \pkg{Pathogen.jl}, which leverages key features of the \proglang{Julia} language. We  provide an example using \pkg{Pathogen.jl} to simulate an epidemic following a susceptible-infectious-removed ($\mathcal{SIR}$) TN-ILM, and then perform inference using observations that were generated from that epidemic. We also demonstrate the functionality of \pkg{Pathogen.jl} with an application of TN-ILMs to data from a measles outbreak that occurred in Hagelloch, Germany in 1861 \citep{1863_pfeilsticker, 1992_oesterle}.
}
\Plainkeywords{Epidemic, individual level models, transmission networks, Bayesian, Julia} 

\Address{
  Justin Angevaare\\
  Department of Mathematics and Statistics\\
  University of Guelph\\
  Guelph, Ontario, Canada\\
  E-mail: \email{jangevaa@uoguelph.ca}\\
  URL: \url{https://jangevaare.github.io}\\
  \\
  Zeny Feng\\
  Department of Mathematics and Statistics\\
  University of Guelph\\
  Guelph, Ontario, Canada\\
  E-mail: \email{zfeng@uoguelph.ca}\\
  URL: \url{https://zfeng.uoguelph.ca}\\
  \\
  Rob Deardon\\
  Faculty of Veterinary Medicine\\
  Department of Mathematics and Statistics\\
  University of Calgary\\
  Calgary, Alberta, Canada\\
  E-mail: \email{robert.deardon@ucalgary.ca}\\
  URL: \url{https://people.ucalgary.ca/~robert.deardon}
}

\begin{document}

\section{Introduction}

\paragraph*{} 
The availability of data to support the use of highly detailed epidemic models is an increasing reality. Individual level models (ILMs) are a framework that accommodates individual specific risk factor information to describe infectious disease dynamics \citep{2010_deardon}. In accounting for population heterogeneity with ILMs, disease dynamics can be captured with more realistic models, and in turn, control strategies can be evaluated using these models. As the number of individuals and the associated data for these individuals increases, the computational requirements for simulation and inference of ILMs are magnified. With epidemic modelling in support of decision making in ongoing epidemics, time, whether for development or computation, is at a premium. Time constraints aside, the epidemiologists that utilize these models may not have the training to support development of a high performance implementation of a specialized ILM in a low level language.

\paragraph*{}
\proglang{Julia} is a high level, high performance language that has been motivated by the needs of modern scientific computing \citep{2018_bezanson}. Previously, the use of lower level compiled languages, such as \proglang{Fortran} or \proglang{C++}, has typically been required to implement high performance features of packages in high level scientific computing languages such as \proglang{R} \citep{2018_bezanson}. This \textit{two language problem} is insidious, and generates an unnecessary divide between those that can use software, and those that can review or advance its development. We use \proglang{Julia} to avoid this problem. As a high level and high performance language, \proglang{Julia} is well situated to address some of the ongoing challenges and opportunities in developing and deploying specialized ILMs.

\paragraph*{}
To date, there have been no packages specific for epidemic modelling published on the \proglang{Julia} General Registry. There are \proglang{Julia} packages however that could be used for this purpose. \pkg{DifferentialEquations.jl} is a comprehensive, general purpose package for differential equations \citep{2017_rackauckas&nie}, and is well suited for population-level epidemic modelling through its discrete stochastic differential equation functionality \citep{2017_rackauckas&nie_b}. Use of these kinds of models can be complementary to the use of ILMs \citep{2017_webb}. \pkg{DifferentialEquations.jl} can simulate from population-level epidemic models, and provides several methods of Bayesian parameter estimation through \pkg{DiffEqBayes.jl}. The \pkg{BioSimulator.jl} package provides simulation methods for interacting populations \citep{2018_landeros}, and could also be used for population level epidemic models. While there are multiple simulation algorithms and full flexibility into the interaction network of the simulated entities in \pkg{BioSimulator.jl}, there are no inference methods provided.

\paragraph*{}
Modelling of heterogeneous populations can be done with \pkg{Agents.jl}, which provides a general purpose package for stochastic simulation of agents on grid systems in discrete time \citep{2019_vahdati}. Agent Based Models (ABMs) can incorporate agent movement and complex behaviours into the epidemics they generate. They prove to be powerful tools in the consideration of individual decision making, especially in the context of learning and behavioural shifts by individuals during an epidemic, and relating those individual actions to higher level disease dynamics, such as seen in \cite{2020_abdulkareem}. Such dynamics are not readily modelled by other methods, but ABMs are also limited in their ability to fit with traditional inference procedures. Approximate Bayesian Computation (ABC) is one avenue for ABM parameterization, where prior beliefs on realistic parameter values are refined using measures of similarity between flexibly summarized ABM simulations and experimental or observational data \citep{2017_ross}. Which is to say, ABMs, like ILMs, and stochastic differential equation models all have their place in epidemics research. 

\paragraph*{}
While not registered on the Julia General Registry at this time, \cite{2020_shoukat} use \proglang{Julia} in the construction of an agent based model specific to the application of predicting hospital and ICU capacity in Canada during the COVID-19 pandemic. Using a population of 10000 agents, they perform simulations to assess different scenarios in regards to resulting proportions of the population requiring care in a hospital or ICU.

\paragraph*{}
Outside of \proglang{Julia}, there are several packages for epidemic modelling on \proglang{R}'s CRAN \citep{R}. Amongst these epidemic modelling packages, there are two for working with ILMs specifically: \pkg{EpiILM} \citep{EpiILM} and \pkg{EpiILMCT} \citep{EpiILMCT}. \pkg{EpiILMCT} is a package for network and spatial continuous time ILMs. This package offers a high performance low level implementation in \proglang{Fortran} for performing MCMC for continuous time ILMs. To maintain performance, functions of risk factors are limited to the form of coefficient and power-parameter. This package offers data augmentation for epidemics in which event times are assumed to be unknown. \pkg{EpiILM}, on the other hand, supports network and spatial ILMs in discrete time, and does not have data augmentation functionality. \pkg{EpiILM} similarly has its core functionality programmed in \proglang{Fortran} for performance reasons.

\paragraph*{}
Beyond the \proglang{R} epidemic modelling packages that utilize the ILM framework of \cite{2010_deardon}, there is also \pkg{epinet} which provides tools for modelling epidemics that spread over contact networks described by exponential-family random graph models \citep{2018_Groendyke&Welch}. This capable package can perform simulations and inference with optional event time augmentation. Its core functionality is implemented in \proglang{C}. Another is \pkg{EpiModel}, which provides functionality to build and simulate individual based epidemic models over stochastic and dynamic contact networks \citep{2018_jenness}. The \proglang{R} package \pkg{surveillance} is also notable for its implementation of several spatio-temporal epidemic models, including individual level epidemic models \citep{2017_meyer}. However, the individual level models in \pkg{surveillance} do not impute transmission networks or event times.

\paragraph*{}
The remainder of the paper is organized as follows. In Section \ref{ilms}, we describe ILMs \citep{2010_deardon}, and then an extension of ILMs, the Transmission Network Individual Level Model (TN-ILM). Section \ref{methods} presents methods for simulating epidemics and fitting TN-ILMs. In Section \ref{julia_implementation}, we introduce and present in detail, our high performance package, \pkg{Pathogen.jl}, for implementing the simulation, model fitting, and inference of TN-ILMs with data augmentation capability in continuous time. Section \ref{ex} demonstrates how to use \pkg{Pathogen.jl} to simulate epidemics and how to fit TN-ILMs via MCMC to simulated data sets through an example. An application to the measles epidemic in Hagelloch, Germany in 1861 \citep{1863_pfeilsticker, 1992_oesterle} is then provided in Section \ref{1861application}. The paper concludes in Section \ref{future} with a brief discussion of future work.

\section{Individual level models}\label{ilms}
\paragraph*{} 
In an ILM, each individual is considered to be in one of several disease states at any time, and the rates governing their transition through the disease states are a function of both individual specific risk factors and the disease states of other individuals in the population at that time. As a whole, the disease state transitions, and subsequent disease state transition rate updates constitute a time-heterogeneous Poisson process.

\paragraph*{}
ILMs may be continuous or discrete with respect to time. In continuous time ILMs, we think of the temporal data structure as consisting of variable time periods, and these representing the length of time between events; \textit{i.e.}, inter-event periods. Here, events are considered to be the disease state transitions in the population. A discrete time ILM approximates its continuous time counterpart, and generally uses time periods that are equal in duration. The likelihoods and strategies for inference for these models are different, however the disease state transition rates each uses can be defined in the same way.

\paragraph*{}
There is flexibility in the disease states that are considered, with susceptible and infectious states being the only necessary states. In the following, the structure of a more complex framework, the susceptible-exposed-infectious-removed ($\mathcal{SEIR}$) ILM, will be described. 

\paragraph*{}
In an $\mathcal{SEIR}$ ILM, the rate that an individual, $i$, transitions from the susceptible state to the exposed state during the $t^{th}$ time period is given as:

\begin{align}
\lambda_{SE}(i, t) = & \left[\Omega_{S}(i)\sum_{k \in \mathcal{I}_{(t)}} \Omega_{T}(k) \kappa(i, k)\right] + \epsilon(i) \text{ for } i \in \mathcal{S}_{(t)},\label{ilm_se}
\end{align}
where,
	\begin{itemize}
	\item $\mathcal{I}_{(t)}$ is the set of infectious individuals during the $t^{th}$ time period,
	\item $\mathcal{S}_{(t)}$ is the set of susceptible individuals during the $t^{th}$ time period,
	\item $\Omega_{S}(i)$ is a function of risk factors associated with the risk of susceptible individual $i$ contracting the disease (susceptibility),
	\item $\Omega_{T}(k)$ is a function of risk factors associated with the risk of infection transmission from the $k^{th}$ individual (transmissibility),
	\item $\kappa(i, k)$ is an infection kernel, a function of risk factors involving both the $i^{th}$ and $k^{th}$ individuals, which often describes the connectivity between these individuals, and,
	\item $\epsilon(i)$ is a function of risk factors associated with exposure to the $i^{th}$ individual that the model otherwise fails to explain. Typically this refers to exposure from a non-specified source outside of the observed population. This is also referred to as the \textit{sparks function}.
	\end{itemize}

\noindent Transition between exposed and infectious states for the $j^{th}$ individual during the $t^{th}$ time period occurs with rate:

\begin{align}
\lambda_{EI}(j, t) = & \Omega_{L}(j) \text{ for } j \in \mathcal{E}_{(t)} \label{ilm_ei}
\end{align}
where,
	\begin{itemize}
	\item $\mathcal{E}_{(t)}$ is the set of exposed individuals during the $t^{th}$ time period, and
	\item $\Omega_{L}(j)$ is a function of risk factors associated with the latent period, the length of time between exposure and onset of infectious in the $j^{th}$ individual.
	\end{itemize}

\noindent Lastly, the $k^{th}$ individual transitions from the infectious to the removed state during the $t^{th}$ time period with rate:

\begin{align}
\lambda_{IR}(k, t) = & \Omega_{R}(k) \text{ for } k \in \mathcal{I}_{(t)} \label{ilm_ir}
\end{align}
where,
	\begin{itemize}
	\item $\Omega_{R}(k)$ is a function of risk factors associated with the removal of the $k^{th}$ individual from an infectious state. Removal can refer to recovery with acquired immunity, death, quarantine, \textit{etc}.
	\end{itemize}
\paragraph*{} In combination, the rates $\lambda_{SE}(i, t)$, $\lambda_{EI}(j, t)$, and $\lambda_{IR}(k, t)$ describe the spread of disease in an $\mathcal{SEIR}$ ILM. 

\paragraph*{}
In a continuous time $\mathcal{SEIR}$ ILM, no change occurs to the sets $\mathcal{S}_{(t)}, \mathcal{E}_{(t)}, \mathcal{I}_{(t)}$, and $\mathcal{R}_{(t)}$ during an inter-event period. It follows that the rates $\lambda_{SE}(i, t)$, $\lambda_{EI}(j, t)$, and $\lambda_{IR}(k, t)$ remain constant during each of these time periods, and that the occurrence of any single event (disease state transition) marks the end of the $t^{th}$ inter-event time period. Movement of an individual into or out of a modelled population may be represented through a deterministic change of state membership of that individual to or from $\mathcal{R}_{(t)}$, followed by the appropriate rate updates. For the remainder of this paper we focus exclusively on continuous time ILMs.

\subsection{Transmission Network ILM extension}\label{TN}
\paragraph*{}
In ILMs, a susceptible individual's risk of being infected by an infectious disease is based on the culmination of various risk factors, such that the influence of specific sources of exposure are masked. We introduce Transmission Network ILMs (TN-ILMs), that instead are explicit with respect to exposure sources. In an $\mathcal{SEIR}$ TN-ILM, a set of competing transition rates for each susceptible individual, $i$, to the exposed state are defined as

\begin{align}
\lambda_{SE}^{*}(i, k, t) =& \Omega_{S}(i)\Omega_{T}(k)\kappa(i, k) \enskip \text{for} \enskip i \in \mathcal{S}_{(t)}, \enskip k \in \mathcal{I}_{(t)} \label{eq-tn-endo}
\end{align}

\noindent describing transition rates specific to each infectious individual, $k$, and with,

\begin{align}
\lambda_{SE}^{*}(i, t) =& \epsilon^{*}(i) \text{ for } i \in \mathcal{S}_{(t)} \label{eq-tn-exo}
\end{align}

\noindent describing the transition rate specific to any exogenous exposure source. While $\epsilon(i)$ in an ILM is not necessarily specific to exogenous sources, that assumption is made with $\epsilon^{*}(i)$ in TN-ILMs.

\section{Methods} \label{methods}
\paragraph*{}
This section starts with a description of simulation methods for continuous time ILMs and TN-ILMs in Subsection \ref{method-CT-ILM-simulation}. In Subsection \ref{sec_ilmllikelihood}, the TN-ILM likelihood function is presented, again drawing comparisons to continuous time ILMs. Lastly, in Subsection \ref{subsec_mcmc}, an approach is detailed for Bayesian inference of TN-ILMs via Markov Chain Monte Carlo (MCMC).

\subsection{Continuous time-to-event simulation} \label{method-CT-ILM-simulation}
\paragraph*{}
As a Poisson process, an epidemic following a continuous time ILM can be stochastically simulated using the \cite{1977_gillespie} algorithm. Here, the inter-event period is the minimum of competing exponential random variables, with each of these exponential random variables representing a possible event given the disease status of the population. The minimum, and the inter-event time, is generated directly from an exponential distribution with rate

\begin{align}
\upsilon(t) =& \sum_{i \in \mathcal{S}_{(t)}} \lambda_{SE}(i, t) + \sum_{j \in \mathcal{E}_{(t)}} \lambda_{EI}(j, t) + \sum_{k \in \mathcal{I}_{(t)}} \lambda_{IR}(k, t).
\end{align}

The specific event that occurs at this time is generated from a multinomial distribution with probability vector

\begin{align}
\pi_{1}(t) =& \begin{bmatrix}
\frac{\lambda_{SE}(1, t)}{\upsilon(t)}, &
\hdots, &
\frac{\lambda_{SE}(N, t)}{\upsilon(t)}, &
\frac{\lambda_{EI}(1, t)}{\upsilon(t)}, &
\hdots, &
\frac{\lambda_{EI}(N, t)}{\upsilon(t)}, &
\frac{\lambda_{IR}(1, t)}{\upsilon(t)}, &
\hdots, &
\frac{\lambda_{IR}(N, t)}{\upsilon(t)}
\end{bmatrix}^{\top}, \label{eq_pi1}
\end{align}

\noindent for a population of size $N$. Which is to say, the probability of a specific event, being the first event to occur amongst competing events, is proportional to its contribution to $\upsilon(t)$. If the generated event is the transition of a susceptible individual to the exposed state, and a TN-ILM is used, the transmission source is then generated from a multinomial distribution, with probability vector

\begin{align}
\pi_{2}(t) =& \begin{bmatrix} 
\frac{\lambda_{SE}^{*}(i, 1, t)}{\lambda_{SE}(i, t)}, & 
\hdots, &
\frac{\lambda_{SE}^{*}(i, N, t)}{\lambda_{SE}(i, t)}, &
\frac{\lambda_{SE}^{*}(i, t)}{\lambda_{SE}(i, t)}
\end{bmatrix}^{\top} \label{eq_pi2},
\end{align}

\noindent following the same logic. With the generation of an inter-event interval, and generation of the specific event occurrence, the population is then updated (\textit{i.e.}, membership in the sets of $\mathcal{S}_{(t+1)}$, $\mathcal{E}_{(t+1)}$, $\mathcal{I}_{(t+1)}$, and $\mathcal{R}_{(t+1)}$), and the process repeats, until either no further events are possible (\textit{i.e.}, $\pi_{1}(t)$ is a vector of zeros), or until some earlier stop condition is met.

\subsection{Continuous time ILM likelihood} \label{sec_ilmllikelihood}
\paragraph*{}
The likelihood of the continuous time-to-event $\mathcal{SEIR}$ ILM is the product of likelihoods at time periods indexed by $t = 1, \dots T-1$, where $T$ is the total number of events that have occurred. With the length of time since the beginning of the $t^{th}$ time period denoted as $\Delta_{t}$, the likelihood function for the associated parameters, $\bm{\theta}$, is given as:

\begin{align}
L(\bm{\theta}) =& \prod_{t=1}^{T-1} \psi(t) \upsilon(t) \exp \left\{ - \upsilon(t) \Delta_{t} \right\} ,\label{eq_ilm_likelihood}
\end{align}

\begin{align}
\intertext{where,}
\psi(t) =& \begin{cases}
\frac{\lambda_{SE}(i, t)}{\upsilon(t)} \text{ if } i \in (\mathcal{S}_{(t)} \cap \mathcal{E}_{(t+1)}),\text{ \textit{i.e.} } i \text{ transitioned from susceptible to exposed,}\\
\frac{\lambda_{EI}(j, t)}{\upsilon(t)} \text{ if } j \in (\mathcal{E}_{(t)} \cap \mathcal{I}_{(t+1)}), \\
\frac{\lambda_{IR}(k, t)}{\upsilon(t)} \text{ if } k \in (\mathcal{I}_{(t)} \cap \mathcal{R}_{(t+1)}), \end{cases}
\intertext{and,}
N =& \lvert \mathcal{S}_{(t)} \lvert + \lvert \mathcal{E}_{(t)} \lvert + \lvert \mathcal{I}_{(t)} \lvert + \lvert \mathcal{R}_{(t)} \lvert \quad \forall t.
\end{align}

Modification is required for TN-ILMs to account for specific transmissions sources, with the TN-ILM, the likelihood function is given as:

\begin{align}
L(\bm{\theta}) =& \prod_{t=1}^{T-1} \psi^{*}(t) \upsilon(t) \exp \left\{ - \upsilon(t) \Delta_{t} \right\}, \label{eq_tn_ilm_likelihood}
\intertext{where,}
\psi^{*}(t) =& \begin{cases}
\frac{\lambda_{SE}^{*}(i, k, t)}{\upsilon(t)} \text{ if } i \in (\mathcal{S}_{(t)} \cap \mathcal{E}_{(t+1)}) \text{ by endogenous exposure from individual } k,\\
\frac{\lambda_{SE}^{*}(i, t)}{\upsilon(t)} \text{ if } i \in (\mathcal{S}_{(t)} \cap \mathcal{E}_{(t+1)}) \text{ by exogenous exposure},\\
\frac{\lambda_{EI}(j, t)}{\upsilon(t)} \text{ if } j \in (\mathcal{E}_{(t)} \cap \mathcal{I}_{(t+1)}), \\
\frac{\lambda_{IR}(k, t)}{\upsilon(t)} \text{ if } k \in (\mathcal{I}_{(t)} \cap \mathcal{R}_{(t+1)}).
\end{cases}
\end{align}

This likelihood mirrors the simulation process described in Section \ref{method-CT-ILM-simulation}. That is, inter-event periods follow exponential distributions, which are paired with multinomial distributions for the occurrence of specific events. The product of the likelihoods of these distributions, compose the likelihood of the epidemic as a whole.

\subsection{Bayesian inference via Markov chain Monte Carlo} \label{subsec_mcmc}
\paragraph*{}
In the Bayesian framework, beliefs about a parameter are described by a posterior distribution. The posterior distribution is a probability distribution for model parameter values, $\theta$, conditioned on observational data $D$. From a posterior distribution, credible intervals and point estimates for parameters can be obtained, and hypothesis testing can be conducted. Monte Carlo methods, such as Markov chain Monte Carlo (MCMC) methods are typically used to generate a sufficient quantity of samples from the posterior distribution, such that the properties of the posterior distribution can be estimated through the generated samples.

\paragraph*{}
The Metropolis-Hastings (M-H) MCMC algorithm is used to perform Bayesian parameter estimation for continuous time ILMs. The M-H algorithm generates a Markov chain consisting of a sequence of samples within the parameter space, the distribution of which converges to the target distribution - the posterior distribution of the parameters in this case. To generate such a Markov chain, some initial values, here denoted as $\bm{\theta}_{1}$, must be first selected or generated. Following initialization of the Markov chain, new samples are proposed with a transition kernel. For a symmetric transition kernel, a sample proposed for the $w^{th}$ iteration, ${\bm{\theta}_{w}}'$, is accepted with probability
\begin{align}
\alpha(\bm{\theta}_{w}') = P(\bm{\theta}_{w} = \bm{\theta}_{w}') =& \min \left( 1, \dfrac{L(\bm{\theta}_{w}') P(\bm{\theta}_{w}')}{L(\bm{\theta}_{w-1}) P(\bm{\theta}_{w-1})} \right). \label{acceptanceprob}
\end{align}
If the proposal $\bm{\theta}_{w+1}'$ is rejected, $\bm{\theta}_{w}$ remains at $\bm{\theta}_{w-1}$ \citep{1970_hastings, 2013_robert&casella}. 

\paragraph*{}
Sampling from the posterior distribution of continuous time ILMs using the M-H algorithm is straightforward when an epidemic is fully observed; \textit{i.e.}, with event times that are known exactly with certainty. While for certain epidemics complete or nearly complete case identification may occur, exact event times are rarely known. Here, unknown event times are imputed using a \textit{data augmentation} process, in which they are treated as additional parameters to be estimated. This tends to result in a massive increase in the dimensionality of the parameter space, and with that computational challenges arise. With TN-ILMs these challenges are exacerbated by the transmission network, which is latent and requires imputation. A compatibility must be enforced in the generation of MCMC proposals between the transmission network and event times. Without this incorporated into the proposal mechanism, the rejection rate using M-H would be unreasonably high (\textit{i.e.} we must avoid transmission network proposals that are impossible given the order of infection of individuals, as dictated by the event time set). As such, a specialized MCMC algorithm is required for inference of TN-ILMs.

\subsubsection*{Initialization strategy} \label{initialization}
\paragraph*{}
Initialization is an important step to facilitate efficient MCMC in these models. High dimensionality and event time interdependence results in vast areas of the parameter space having near-zero posterior density, which is approximated to zero in computation. With MCMC, it may take a long time to move into and sample from areas of the parameter space that have non-zero computationally approximated posterior mass. This also impacts adaptive tuning of transition kernel variance and further contributes to bottlenecks in MCMC convergence. Our strategy is to generate many potential sets of initial values, and to select the set with the highest posterior density. Multiple Markov chains, and a higher number of initialization generations per Markov chain are recommended for higher dimensional applications of TN-ILMs, as is common when implementing MCMC in general \citep{2013_robert&casella}. These independently initialized Markov chains allow for a more substantiated assessment of Markov chain convergence, in comparison to assessing the behaviour of a single Markov chain in isolation.

\paragraph*{}
The initialization of each Markov chain begins with sampling a set of model parameters from their corresponding prior distributions. A set of event times is also generated for observation delays, as well as latent periods if applicable. Since the transmission network can be generated directly from its conditional distribution with TN-ILMs, we marginalize the TN-ILM likelihood over all possible transmission networks, simplifying it to the standard ILM likelihood for the purposes of Markov chain initialization. The ILM likelihood is calculated by iterating through each event in the generated set of event times. While this calculation is computationally intensive, conveniently, it can be stopped early when its running result drops below some threshold. This convenience is leveraged in our initialization generation process such that computational cost does not increase linearly with an increase in the number of initialization attempts.

\subsubsection*{Iteration strategy}
\paragraph*{}
Each MCMC iteration can be broken down into several sub-steps: event time sampling, parameter sampling, and transmission network sampling. 

\paragraph*{Event times}
A random walk Metropolis-Hastings sampling procedure is used to sample event times from the TN-ILM posterior distribution. Specifically a bounded normal distribution is used as the transition kernel, with its mean set to the event time in the previous iteration, and with a pre-specified variance. These bounds are determined by the times of the events that are dependent through the transmission network or model structure. 

\paragraph*{}
The event times relevant to determining the bounds of event time proposal distributions include: the times of other state transitions by the involved individual (\textit{e.g.}, an individual may not transition to a removed state prior to being in an infected state); the infection and, if applicable, removal times of their transmission source (\textit{e.g.}, a susceptible individual may only transition to the exposed state while their transmission source is in an infected state). Similarly, if an individual transmits the disease themselves, they must be in the infected state prior to their earliest transmission, and must not be removed until after their final transmission. This ensures the transmission network remains compatible with the set of event times.

\paragraph*{}
Due to event time interdependence, event time proposals are generated one at a time. The order in which event times are updated is randomized at each iteration. A decision to accept or reject proposed event times can be conducted in batches such that fewer likelihood calculations occur. Successful tuning of the event time batch size and transition kernel variance results in effective sampling of the event time posterior distribution while minimizing computational costs. To achieve this, experimental tuning is required for each TN-ILM application by the user.

\paragraph*{TN-ILM parameters}
Following updates to the event times, a new set of TN-ILM parameters is proposed, and the proposed values are subject to the acceptance rule for M-H MCMC algorithms shown in Equation \ref{acceptanceprob}. For these proposals, a multivariate normal transition kernel is used. The covariance matrix of the transition kernel can be automatically tuned during sampling, using the adaptive sampling method of \cite{2007_roberts}.

\paragraph*{Transmission network}
Finally, the transmission network is updated using the Gibbs sampler. Each transmission source is sampled from a multinomial distribution corresponding to its conditional distribution. The probability vector of each multinomial distribution is generated following Equation \ref{eq_pi2}, for each applicable time period using the current set of event times and model parameters.


\section[Software implementation with Julia]{Software implementation with {\proglang{Julia}}}\label{julia_implementation}
\paragraph*{}
We have implemented simulation and inference methods for TN-ILMs as described in Section \ref{methods} in \proglang{Julia} with the \pkg{Pathogen.jl} package. \pkg{Pathogen.jl} is open source and released with an MIT (Expat) license. Our description of \pkg{Pathogen.jl} is consistent with its \code{v0.4.12} release.

\paragraph*{}
\pkg{Pathogen.jl} leverages several of the features of the \proglang{Julia} Language. First is the ability for the user to define functions for disease state transition rates with a high level of flexibility without sacrificing performance. While \proglang{Julia} is an interactive language, with a familiar Read-Evaluate-Process-Loop (REPL) interface, behind that, \proglang{Julia} code is Just-In-Time (JIT) compiled to highly optimized machine code. In a set of benchmarks performed by \cite{2017_bezanson}, \proglang{Julia} was usually found to be within a factor of two in regards to computation time of equivalent \proglang{C} code.

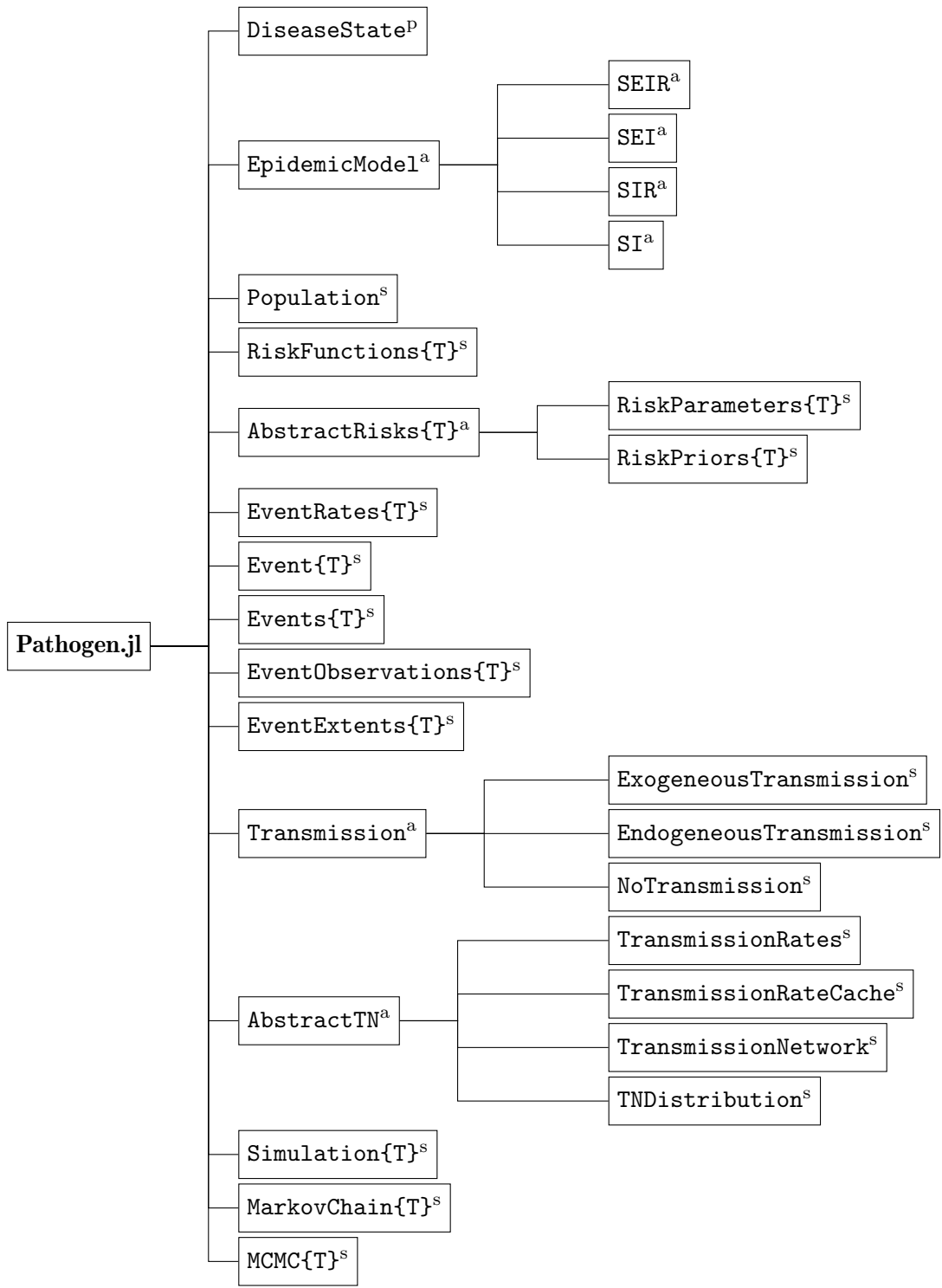
\begin{figure}
\begin{tikzpicture}
\tikzset{grow'=right} 
\tikzset{edge from parent/.style={draw, thin,
    edge from parent path={(\tikzparentnode.east)
        -- +(25pt, 0pt)
        |- (\tikzchildnode.west)}}}
\tikzset{level 1/.style={level distance=100pt}}
\tikzset{level 2/.style={level distance=160pt}}
\tikzset{execute at begin node=\strut} 
\tikzset{every tree node/.style={anchor=west, draw}}

\Tree [.\pkg{Pathogen.jl} 
    [.\code{DiseaseState}\textsuperscript{p} ]
    [.\code{EpidemicModel}\textsuperscript{a} 
        [.\code{SEIR}\textsuperscript{a} ]
        [.\code{SEI}\textsuperscript{a} ]
        [.\code{SIR}\textsuperscript{a} ]
        [.\code{SI}\textsuperscript{a} ] ]
    [.\code{Population}\textsuperscript{s} ]
    [.\code{RiskFunctions\{T\}}\textsuperscript{s} ]
    [.\code{AbstractRisks\{T\}}\textsuperscript{a}
        [.\code{RiskParameters\{T\}}\textsuperscript{s} ]
        [.\code{RiskPriors\{T\}}\textsuperscript{s} ] ]
    [.\code{EventRates\{T\}}\textsuperscript{s} ]
    [.\code{Event\{T\}}\textsuperscript{s} ]
    [.\code{Events\{T\}}\textsuperscript{s} ]
    [.\code{EventObservations\{T\}}\textsuperscript{s} ]
    [.\code{EventExtents\{T\}}\textsuperscript{s} ]
    [.\code{Transmission}\textsuperscript{a} 
        [.\code{ExogeneousTransmission}\textsuperscript{s} ]
        [.\code{EndogeneousTransmission}\textsuperscript{s} ]
        [.\code{NoTransmission}\textsuperscript{s} ] ]
    [.\code{AbstractTN}\textsuperscript{a} 
        [.\code{TransmissionRates}\textsuperscript{s} ]
        [.\code{TransmissionRateCache}\textsuperscript{s} ]
        [.\code{TransmissionNetwork}\textsuperscript{s} ]
        [.\code{TNDistribution}\textsuperscript{s} ] ]
    [.\code{Simulation\{T\}}\textsuperscript{s} ]
    [.\code{MarkovChain\{T\}}\textsuperscript{s} ]
    [.\code{MCMC\{T\}}\textsuperscript{s} ] ]
\end{tikzpicture}
\caption{The hierarchy of the types provided by \pkg{Pathogen.jl} are shown above. Type names followed by \code{\{T\}} indicate that that type is parametric - in all cases the type is parametrized by a subtype of \code{EpidemicModel}. Subscripts are used to indicate the kind of type: \textsuperscript{p}primitive: elementary data representations; \textsuperscript{a}abstract: types used for organization purposes or to invoke specific function methods but do not contain data; \textsuperscript{s}struct: types that have fields containing other types. \label{typetree}}
\end{figure}

\paragraph*{} 
In \proglang{Julia}, new user-defined types are implemented without performance penalty in comparison to the basic types provided by the language. This is made no clearer than by the fact that much of \proglang{Julia}'s features, are written in \proglang{Julia} itself. \proglang{Julia}'s \textit{multiple dispatch} enables both generalist and specialist function methods that are invoked based on argument types. The type system results in code that can often be common across a set of types. \pkg{Pathogen.jl} provides types to support $\mathcal{SEIR}$, $\mathcal{SEI}$, $\mathcal{SIR}$, and $\mathcal{SI}$ TN-ILMs. With only some basic methods defined for these types or across unions of these types, higher level code such as that involved in epidemic simulation, likelihood calculations, and performing MCMC is able to be kept common across all disease model classes. This same functionality in theory would reduce the development time required to implement additional disease model classes. The types implemented by \pkg{Pathogen.jl}, and their hierarchy are shown in Figure \ref{typetree}.

\paragraph*{} Distributed computing functionality is highly relevant, if not essential for large data applications. For TN-ILMs, this would be for modelling large populations and/or higher complexity models. \proglang{Julia} has been designed for simple, but powerful distributed computing. Some of this is leveraged in \pkg{Pathogen.jl} for conducting MCMC for TN-ILMs. Independent Markov chains are easily initialized, and ran across multiple cores, or on a high performance cluster in parallel.

\subsection[Installing Pathogen.jl]{Installing {\pkg{Pathogen.jl}}}
\paragraph*{} \pkg{Pathogen.jl} \code{v0.4.12} is available for \proglang{Julia} version 1.1 and higher, and is listed on \proglang{Julia}'s General Registry. Julia has a built-in package manager, \pkg{Pkg}. \pkg{Pkg} provides a special REPL that is entered from the Julia REPL by typing a single \code{]}. From the package REPL, \pkg{Pathogen.jl} \code{v0.4.12} is installed with \code{add Pathogen@v0.4.12}. Users also have the option to install \pkg{Plots.jl} to make use of custom plotting tools provided by \pkg{Pathogen.jl}. This can be done with \code{add Plots} from the package REPL. Once installed, \pkg{Pathogen.jl} can be immediately used with \code{using Pathogen} from back in the Julia REPL. This entire process takes only a few seconds.

\subsection[Pathogen.jl basics]{{\pkg{Pathogen.jl}} basics}
\paragraph*{} Whether using \pkg{Pathogen.jl} for simulation or inference of TN-ILM epidemics, a population must be first defined. We represent populations in our package with a  \code{Population} type. More specifically, our \code{Population} type is a \textit{struct}. In \proglang{Julia}, a struct is a type that is composed of other types - perhaps other structs, or \textit{primitive} types such as \code{Float64} or \code{Bool}. 

\paragraph*{} To construct a \code{Population}, a \code{DataFrame} containing individual specific risk factors must be specified. \code{DataFrame} is a type provided by the \pkg{DataFrames.jl} package, which is a dependency of \pkg{Pathogen.jl}. In this \code{DataFrame}, each row will represent an individual. There is also an option to provide a distance matrix when describing a population. Distance measures are common components of infectivity kernels. The distance matrix can be used to avoid repeated calculation of these distances from individual specific risk factor information. \code{Population} in \pkg{Pathogen.jl} is declared as
\\\\
\begin{minipage}{\textwidth}
\begin{Code}
struct Population
  risks::DataFrame
  distances::Union{Nothing, AbstractArray}
  individuals::Int64
end
\end{Code}
\end{minipage}

\paragraph*{} 
The structure of TN-ILMs is described with \code{RiskFunctions\{T\}}, which is a collection of functions that calculate individual specific disease state transition rates. \code{RiskFunctions\{T\}} is a \textit{parametric struct}, declared as shown below, with slightly different construction and behaviour for different values of \code{T}.

\begin{Code}
struct RiskFunctions{T<: EpidemicModel}
  sparks::Union{Nothing, Function}
  susceptibility::Union{Nothing, Function}
  infectivity::Union{Nothing, Function}
  transmissibility::Union{Nothing, Function}
  latency::Union{Nothing, Function}
  removal::Union{Nothing, Function}
end
\end{Code}

\paragraph*{}
Parametric structs are used throughout \pkg{Pathogen.jl} to provide specialization to different disease model classes. This allows for modified functionality where it is needed for the various model class implementations in otherwise common code. For instance, construction of \code{RiskFunctions\{SIR\}} when compared to \code{RiskFunctions\{SEIR\}} does not involve specification of a function describing the transition rate between exposed and infectious classes (\textit{i.e.}, $\Omega_{L}$ in Equation \ref{ilm_ei}). 

\paragraph*{}
There is full flexibility in the form of TN-ILM risk functions used in \code{RiskFunctions\{T\}}, as long as these risk functions follow an expected signature for their arguments. Each risk function must accept a \code{Population}, a parameter vector (\code{Vector{Float64}}), and an \code{Int64} individual identifier, as arguments. Infectivity kernels are an exception to this, and must accept two \code{Int64} identifiers - for infection source and newly infected individual. Each risk function should return a \code{Float64}. A complete example of constructing \code{RiskFunctions\{SIR\}}, including the construction of the risk functions that compose it, is provided in Section \ref{ex}.

\paragraph*{}
The parameterization of TN-ILM risk functions is represented by a separate type,\\ \code{RiskParameters\{T\}}. For each of the risk functions required by a TN-ILM, a parameter vector must be provided in \code{RiskParameters\{T\}}, which has been declared as:

\begin{Code}
struct RiskParameters{T<: EpidemicModel} <: AbstractRisk{T}
  sparks::Union{Nothing, AbstractVector}
  susceptibility::Union{Nothing, AbstractVector}
  infectivity::Union{Nothing, AbstractVector}
  transmissibility::Union{Nothing, AbstractVector}
  latency::Union{Nothing, AbstractVector}
  removal::Union{Nothing, AbstractVector}
end
\end{Code}

\subsection[Simulation with Pathogen.jl]{Simulation with {\pkg{Pathogen.jl}}} \label{pathogen-sim}

\paragraph*{} 
To simplify the simulation interface, we have provided a \code{Simulation\{T\}} struct with \pkg{Pathogen.jl}. \code{Simulation\{T\}}s include all of the information required to iterate and track the progression of an epidemic. Once constructed, the \code{simulate!} function is used to run the simulation, updating event times, individual disease states, and the transmission network within the \code{Simulation\{T\}} as appropriate. The \code{simulate!} function will iterate until a specified stop condition is met (processing time, simulation time, and/or number of iterations) or if there are no further events possible. \code{Simulation\{T\}} is a \textit{parametric mutable struct}, which is like a parametric struct, except it allows for its values to be changed or updated. For instance, the \code{Int64} value for \code{iterations}, shown in the type declaration below, is incremented at each iteration, which is possible because of this mutability. Explicitness about struct mutability in \proglang{Julia} allows for certain optimizations by the compiler when dealing with structs that do not change in composition.
\begin{Code}
mutable struct Simulation{T <: EpidemicModel}
  time::Float64
  iterations::Int64
  population::Population
  risk_functions::RiskFunctions{T}
  risk_parameters::RiskParameters{T}
  disease_states::Vector{DiseaseState}
  transmission_rates::TransmissionRates
  event_rates::EventRates{T}
  events::Events{T}
  transmission_network::TransmissionNetwork
end
\end{Code}
The use of \code{\{T\}} throughout the declaration of \code{Simulation\{T\}} ensures matching type parameterizations for all of the types that compose it. This also means that the parameterization of \code{Simulation\{T\}} can be inferred during its construction.
\paragraph*{}
There are several ways to construct a \code{Simulation\{T\}}. The simplest construction method requires only specification of a \code{Population}, \code{RiskFunctions\{T\}}, and \code{RiskParameters\{T\}} for the TN-ILM. In this case, an entirely susceptible population is assumed for a starting time of $0.0$ time units. The internal code of this basic \code{Simulation\{T\}} construction method is shown below. Individuals can start the simulation from other specified \code{DiseaseState}s and/or the simulation may have a different start time using the other construction methods (not shown).

\begin{minipage}{\textwidth}
\begin{Code}
  function Simulation(pop::Population,
                      rf::RiskFunctions{T},
                      rp::RiskParameters{T}) where T <: EpidemicModel
    states = fill(State_S, pop.individuals)
    tr = initialize(TransmissionRates, states, pop, rf, rp)
    rates = initialize(EventRates, tr, states, pop, rf, rp)
    events = Events{T}(pop.individuals)
    net = TransmissionNetwork(pop.individuals)
    return new{T}(0.0, 0, pop, rf, rp, states, tr, rates, events, net)
  end
\end{Code}
\end{minipage}

\paragraph*{} 
Finally, observational data can be generated from a completed \code{Simulation} using the provided \code{observe()} function, with statistical distributions specified for observation delays, in the form of \code{UnivariateDistribution}s from the \pkg{Distributions.jl} package \citep{2019_besancon}, another dependency of \pkg{Pathogen.jl}. 

\subsection[Inference with Pathogen.jl]{Inference with {\pkg{Pathogen.jl}}} \label{pathogen-inf}
\paragraph*{}
As with simulation in \pkg{Pathogen.jl}, performing parameter estimation for TN-ILMs via MCMC requires the definition of a \code{Population} and of \code{RiskFunctions\{T\}}. \code{RiskParameters} are now unknown, and will be sampled from the posterior distribution through MCMC. In order to do this, specification of prior distributions for each risk parameter is first required. These priors are structured through the \code{RiskPriors\{T\}} type. This parametric struct has the same form as \code{RiskParameters}, but in place of each parameter value, a \code{UnivariateDistribution} must be provided.

\paragraph*{}
We also must specify priors for the event time data augmentation process. Currently Uniform prior distributions are supported and applied broadly through an \code{EventExtents\{T\}} struct. These are upper bounds on the length of time between observations of infectiousness, and the actual onset of infection, as well as the length of time between removal observations and actual removal times. For TN-ILMs that have an exposed class, a bound on the length of the latent period is also specified in use of \code{EventExtents\{T\}}.

\begin{Code}
struct EventExtents{T <: EpidemicModel}
  exposure::Union{Nothing, Tuple{Float64, Float64}}
  infection::Union{Nothing, Tuple{Float64, Float64}}
  removal::Union{Nothing, Tuple{Float64, Float64}}
end
\end{Code}

\paragraph*{}
Finally, an \code{MCMC\{T\}} struct is constructed, containing all of the information required for performing MCMC, common across individual Markov chains (\textit{e.g.}, observational data, population data, prior distributions), as well as the individual Markov chains that sample from the TN-ILM posterior distribution. The composition of \code{MCMC\{T\}} is:
\\\\
\begin{minipage}{\textwidth}
\begin{Code}
mutable struct MCMC{T <: EpidemicModel}
  event_observations::EventObservations{T}
  event_extents::EventExtents{T}
  population::Population
  starting_states::Vector{DiseaseState}
  risk_functions::RiskFunctions{T}
  risk_priors::RiskPriors{T}
  transmission_network_prior::Union{Nothing, TNDistribution}
  markov_chains::Vector{MarkovChain{T}}
end
\end{Code}
\end{minipage}
\\
\\
\\
\code{MarkovChain\{T\}} includes vectors of event times from data augmentation, networks, as well as TN-ILM parameters, and is declared as:
\begin{Code}
mutable struct MarkovChain{T <: EpidemicModel}
  iterations::Int64
  events::Vector{Events{T}}
  transmission_network::Vector{TransmissionNetwork}
  risk_parameters::Vector{RiskParameters{T}}
  log_posterior::Vector{Float64}
  cov::OnlineStats.CovMatrix
end
\end{Code}

\paragraph*{}
To construct an \code{MCMC\{T\}} object to perform inference, \code{EventObservations\{T\}}, \code{EventExtents\{T\}}, \code{Population}, \code{RiskFunctions\{T\}}, and \code{RiskPriors\{T\}} must all be specified. Specification of a \code{TNDistribution} as a prior distribution for the transmission network is optional. If a prior for the transmission network is not specified, a flat uniform prior is used. After construction of an \code{MCMC} object, it is initialized using the \code{start!} function. During initialization the user must specify the number of chains to initialize, and the number of initialization attempts per chain. Each chain can be initialized on different cores. After initialization, MCMC can proceed using the \code{iterate!} function. With this function the number iterations are specified, as well as transition kernel variance for event time data augmentation. MCMC for each \code{MarkovChain\{T\}} can also be ran in parallel. An example of the process of performing MCMC with \pkg{Pathogen.jl} is provided in Section \ref{ex}.

\section{Simulated Example}\label{ex}
\paragraph*{}
In the following we present a full example using \pkg{Pathogen.jl} to:
\begin{itemize}
\item Generate an epidemic population,
\item Simulate from an $\mathcal{SIR}$ TN-ILM,
\item Simulate observations from the epidemic, and, 
\item Use the observations to estimate event times, transmission network, and TN-ILM parameters via MCMC.
\end{itemize}
The source code to replicate this example exactly is included in \code{SIR_simulation.jl}.

\paragraph{} We start by loading the various publicly available \proglang{Julia} packages used in the example, and setting the seed of the random number generator:
\begin{Code}
using Distances,
      Pathogen,
      Random,
      Plots

Random.seed!(11235);
\end{Code}
\paragraph*{}
We generate risk factor data for a population containing 100 individuals. A location ($x$ and $y$ coordinates over a $15 \times 30$ unit area), and an arbitrary $\text{Gamma}(\alpha = 1, \beta = 1)$ distributed risk factor:
\begin{Code}
n = 100
risks = DataFrame(x = rand(Uniform(0, 15), n),
                  y = rand(Uniform(0, 30), n),
                  riskfactor1 = rand(Gamma(), n))
\end{Code}
\paragraph*{}
We pre-calculate Euclidean distances between individuals in a distance matrix, which is used in the specification of a \code{Population} object:
\begin{Code}
dists = [euclidean([risks[i, :x]; 
                    risks[i, :y]], 
                   [risks[j, :x]; 
                    risks[j, :y]]) for i = 1:n, j = 1:n]
pop = Population(risks, dists)
\end{Code}
\paragraph*{}
Next, several functions of risk factors are defined with the signature expected by \pkg{Pathogen.jl}, and these structured into a \code{RiskFunctions\{SIR\}} object:
\\
\\
\begin{minipage}{\textwidth}
\begin{Code}
function _constant(params::Vector{Float64}, pop::Population, i::Int64)
  return params[1]
end

function _one(params::Vector{Float64}, pop::Population, i::Int64)
  return 1.0
end

function _linear(params::Vector{Float64}, pop::Population, i::Int64)
  return params[1] * pop.risks[i, :riskfactor1]
end

function _powerlaw(params::Vector{Float64}, pop::Population, 
                   i::Int64, k::Int64)
  beta = params[1]
  d = pop.distances[k, i]
  return d^(-beta)
end
\end{Code}
\end{minipage}

\begin{figure}
    \centering
    \includegraphics[width=\textwidth]{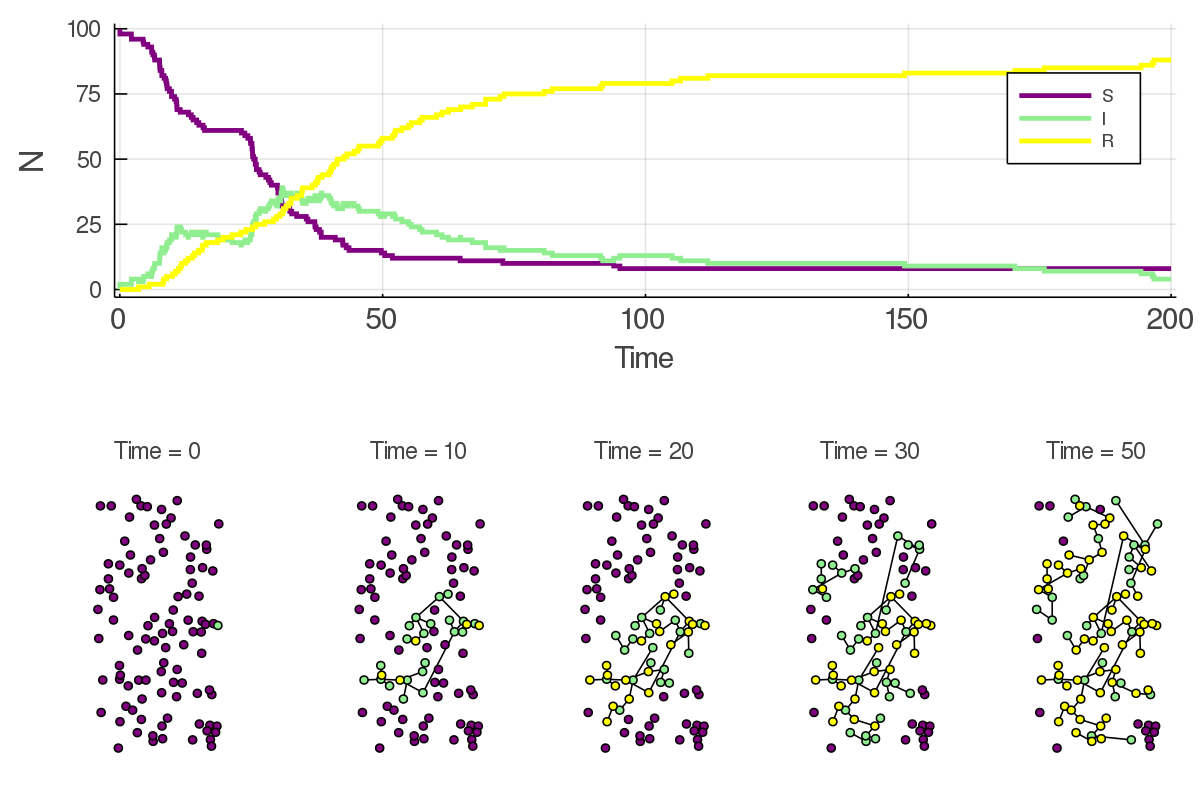}
    \caption{An epidemic simulated from a TN-ILM following code provided in Section \ref{ex}. The top plot shows the overall number of individuals in each of the represented disease states over the length of the epidemic. The bottom row of plots shows individual disease state information along with the state of the transmission network at 5 time points in the epidemic.}
    \label{epiplot}
\end{figure}

\begin{Code}
rf = RiskFunctions{SIR}(_constant, # sparks function
                        _one, # susceptibility function
                        _powerlaw, # infectivity kernel
                        _one, # transmissibility function
                        _linear) # removal function
\end{Code}
These risk functions are then parametrized:
\begin{Code}
rparams = RiskParameters{SIR}([0.0001], # sparks
                              Float64[], # susceptibility
                              [4.0], # infectivity
                              Float64[], # transmissibility
                              [0.1]) # removal
\end{Code}
\paragraph*{} Starting states for each of the 100 individuals are then set, using the \code{DiseaseState} primitive type that is provided by \pkg{Pathogen.jl}. \code{State_S}, \code{State_I}, and \code{State_R} are valid states for the $\mathcal{SIR}$ model. At the start of the epidemic, we'll have individual 1 infectious and the remaining 99 individuals susceptible. The starting states and risk factors of the population, in addition to the parameterized risk functions of the TN-ILM are organized into a \code{Simulation} struct. With this, the simulation is finally ran. The simulation ends once $200$ time units have elapsed in the epidemic.
\begin{Code}
starting_states = append!([State_I], fill(State_S, n-1))
sim = Simulation(pop, starting_states, rf, rparams)
simulate!(sim, tmax=200.0)
\end{Code}
The simulated epidemic can be explored visually. For our example, plots are generated in \proglang{Julia} using \pkg{Plots.jl} \citep{plots} with the \pkg{GR} plotting backend \citep{GR}, with \textit{plot recipes} provided in \pkg{Pathogen.jl}. An epidemic curve can be generated from an \code{Events\{T\}} object, such as the one within our \code{Simulation\{T\}}, with
\begin{Code}
p1 = plot(sim.events)
\end{Code}
and transmission network plots showing the disease states of individuals at specified times, with
\begin{Code}
p2=plot(sim.transmission_network, 
        sim.population, 
        sim.events, 
        0.0, title="Time = 0")
\end{Code}
\dots
\begin{Code}
p6=plot(sim.transmission_network, 
        sim.population, 
        sim.events, 
        50.0, title="Time = 50")
\end{Code}
These plots can then be combined in a layout, to obtain Figure \ref{epiplot} with
\begin{Code}
l = @layout [a;
             b c d e f]

plot(p1, p2, p3, p4, p5, p6, layout=l)
\end{Code}

\begin{figure}
    \centering
    \includegraphics[width=\textwidth]{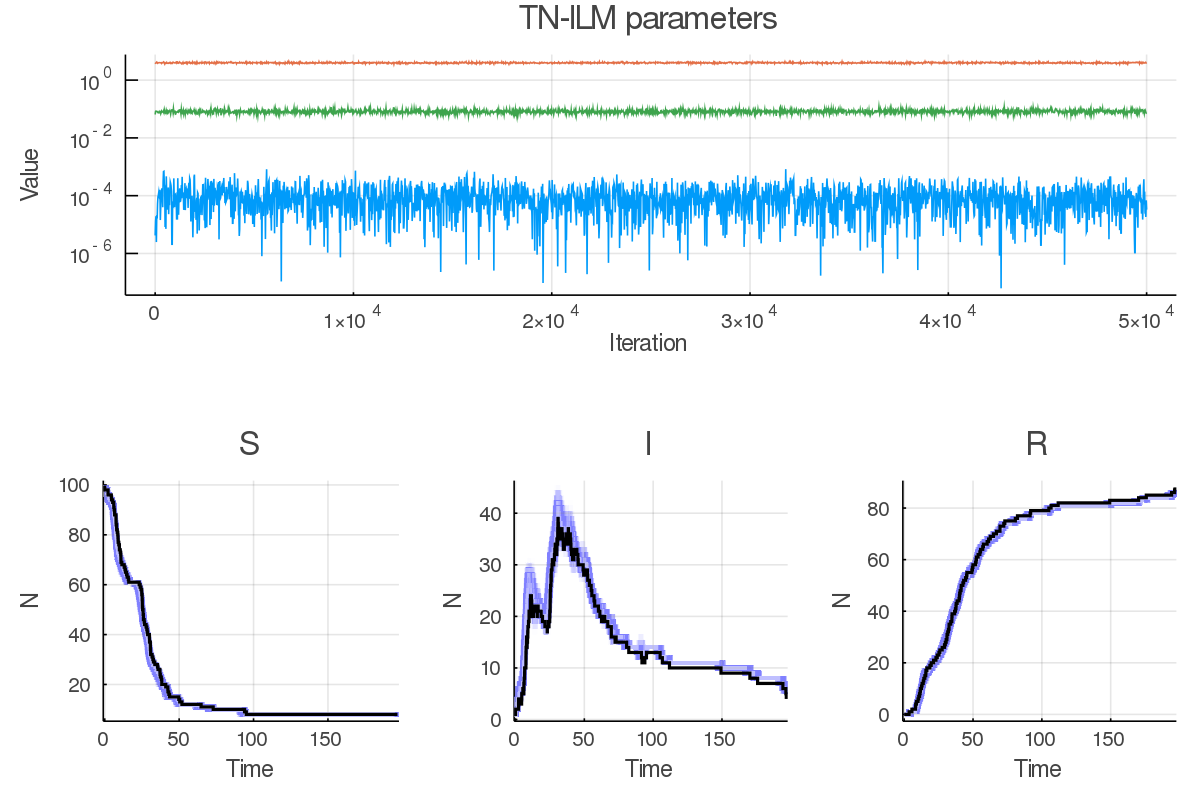}
    \caption{Here we show several plots for visualizing MCMC for a TN-ILM. The top plot shows the values of 3 TN-ILM parameters over 50k iterations. The bottom row of plots shows the epidemic curve posterior distributions in blue (posterior density estimated with every 20th iteration from iteration number 10k to 50k), with the true epidemic curves indicated with black.}
    \label{posteriorplots}
\end{figure}

\begin{figure}
    \centering
    \includegraphics[width=\textwidth]{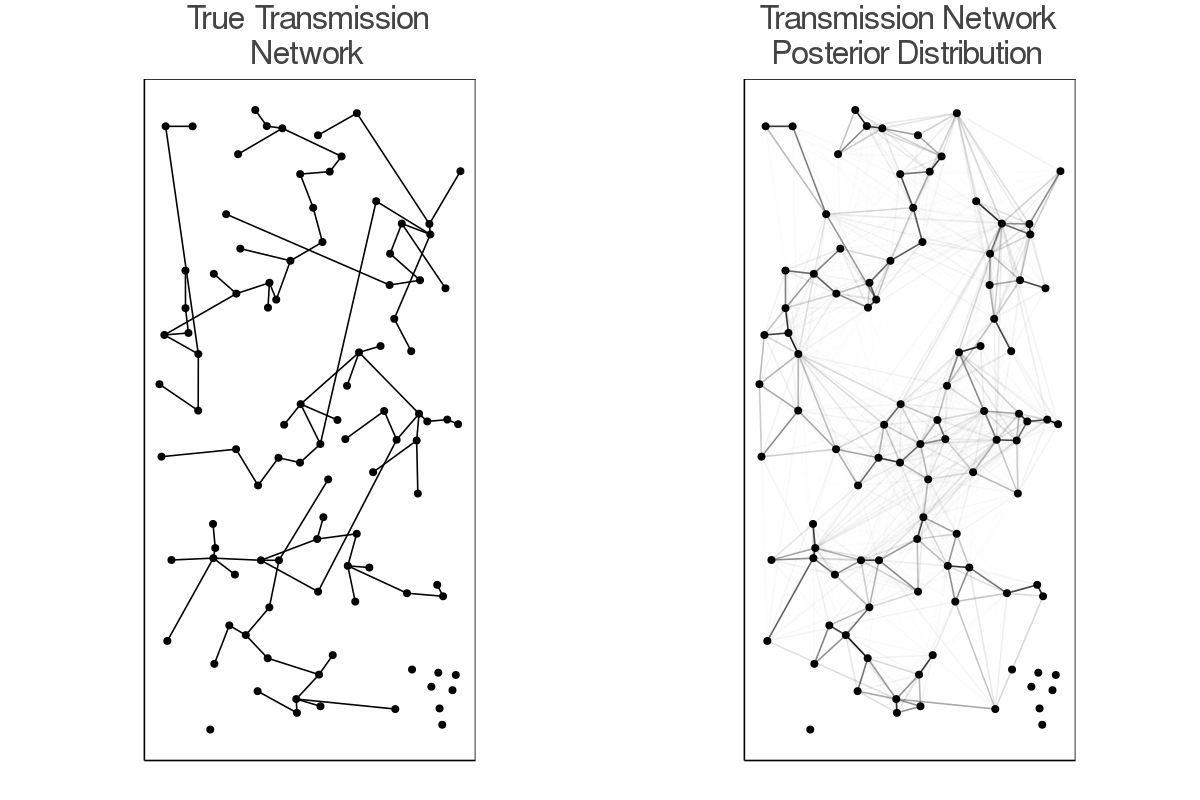}
    \caption{The true transmission network (left), in comparison to the transmission network posterior distribution (right), where the transparency of transmission pathways represents their posterior density.}
    \label{posteriortnplot}
\end{figure}

\paragraph*{}
From the simulated epidemic, observations can be generated, with statistical distributions used for the generation of observation delay. The optional \code{force=true} keyword argument is used to bound the infection observation delay such that an infection observation is guaranteed (\textit{i.e.}, an individual can't move to the removed state undetected).
\begin{Code}
obs = observe(sim, Uniform(0.5, 2.5), Uniform(0.5, 2.5), force=true)
\end{Code}
With observational data, inference for a specified model can be conducted. For our example, we will assume the model structure is known, and reuse the set of risk functions we declared for the epidemic simulation. For each parameter value, a prior distribution must be specified before commencing MCMC. We must also specify priors for our event times through \code{EventExtents}. 
\begin{Code}
rpriors = RiskPriors{SIR}([Exponential(0.0001)],
                          UnivariateDistribution[],
                          [Uniform(1.0, 7.0)],
                          UnivariateDistribution[],
                          [Uniform(0.0, 1.0)])
                           
ee = EventExtents{SIR}(5.0, 5.0)
\end{Code}
MCMC will now be initialized, following the initialization strategy detailed in Section \ref{initialization}, with:
\begin{Code}
mcmc = MCMC(obs, ee, pop, rf, rpriors)
start!(mcmc, attempts=50000) 
\end{Code}
We then perform 50k iterations (second positional argument) for the initialized Markov chain using the \code{iterate!} function. Here, we elect to batch event time data augmentation into 10 sets (specified with the  \code{event_batches} keyword argument), with a transition kernel variance of 1.0 for each event time (third positional argument). We also condition event time augmentation on the previous transmission network (specified with the \code{condition_on_network} keyword argument). 50k iterations in this manner require approximately 18.5 minutes on a computer with an Intel 2.7 GHz i7-3740QM processor, using \proglang{Julia} 1.4.2. For applications to real world data, or for simulations with real world application, longer runs by multiple chains is advised in order to validate convergence.
\begin{Code}
iterate!(mcmc, 50000, 1.0, condition_on_network=true, event_batches=10)
\end{Code}
\paragraph*{} We provide convenient plotting functions for visualizing MCMC and posterior distributions, after MCMC is complete, yielding the plots seen in Figures \ref{posteriorplots} and \ref{posteriortnplot}. The trace plot in Figure \ref{posteriorplots} is generated with:
\begin{Code}
p1 = plot(1:20:50001, mcmc.markov_chains[1].risk_parameters, 
yscale=:log10, title="TN-ILM parameters")
\end{Code}

For the remainder of this example we take every 20th iteration from iteration 10000 through 50000 as being representative samples from the TN-ILM posterior distribution. With this, the epidemic curve posterior distributions are visualized with
\begin{Code}
p2 = plot(mcmc.markov_chains[1].events[10000], 
    State_S, linealpha=0.01, title="S")
for i=10020:20:50000
  plot!(p2, mcmc.markov_chains[1].events[i], State_S, linealpha=0.01)
end
plot!(p2, sim.events, State_S, linecolor=:black)
\end{Code}
for each state. Partially transparent epidemic curves are repeatedly plotted to show the posterior density. We also overlay the true event times in black. A plot layout is specified, and the plots are combined to form Figure \ref{posteriorplots}.
\begin{Code}
l = @layout [a; [b c d]]
plot(p1, p2, p3, p4, layout=l)
\end{Code}

\noindent The \code{TransmissionNetworkDistribution}, or \code{TNDistribution} type for short, is used to represent prior and posterior distributions of Transmission Networks. A plotting method for \code{TNDistribution} is provided, as is exemplified in Figure \ref{posteriortnplot} in comparison to the true transmission network in our simulated epidemic. To generate this plot:
\\
\\
\begin{minipage}{\textwidth}
\begin{Code}
p1 = plot(sim.transmission_network, sim.population, 
          title="True Transmission\nNetwork", framestyle=:box)

tnp = TNDistribution(mcmc.markov_chains[1].transmission_network[10000:20:50000])
p2 = plot(tnp, sim.population, 
          title="Transmission Network\nPosterior Distribution", framestyle=:box)

plot(p1, p2, layout=(1, 2))
\end{Code}
\end{minipage}
\\
\\
\\
A \code{summary} function is also provided which outputs a \code{DataFrame} with summary statistics for each TN-ILM parameter. A burn-in period, and a thinning rate can be provided:
\\
\begin{Code}
summary(mcmc, burnin=10000, thin=20)
\end{Code}
In our simulated example, this generates the results in Table \ref{sim_output}. The parameter values that were used to generate the epidemic simulation were $\zeta = 0.0001, \beta = 4.0$, and $\eta = 0.1$. All 3 of these parameters are contained in the 95\% credible intervals.

\begin{table}
\begin{center}
\begin{tabular}{|c|c|c|c|}
\hline
Parameter & Mean & Variance & 95\% Credible Interval\\
\hline
$\zeta$ & $1.033\times10^{-4}$ & $9.861\times10^{-9}$ & $(2.869\times10^{-6}, 3.686\times10^{-4})$ \\
$\beta$ & $3.967$ & $1.794\times10^{-2}$ & $(3.715, 4.241)$ \\
$\eta$ & $8.168\times10^{-2}$ & $8.457\times10^{-5}$ & $(6.520\times10^{-2}, 1.001\times10^{-1})$ \\
\hline
\end{tabular}   
\end{center}
\caption{The \code{summary()} function outputs a \code{DataFrame} containing estimates of posterior mean, posterior variance, as well as credible intervals for each parameter in a TN-ILM, when applied to \code{MCMC} objects.\label{sim_output}}
\end{table}

\section{Application: 1861 Hagelloch measles outbreak} \label{1861application} 
\paragraph*{} We apply the methods of \pkg{Pathogen.jl} to fit an $\mathcal{SEIR}$ TN-ILM to the 1861 Hagelloch measles data set, originally published by \cite{1863_pfeilsticker}. This data set includes detailed observations on 188 children in Hagelloch, Germany. The data were analyzed by \cite{1992_oesterle}, who expanded the individual level detail of the data set, and predicted the transmission source for each case. The 1861 Hagelloch measles data have since been used to demonstrate numerous new analysis methods and new software in the area of individual level epidemic modelling and transmission network models. \cite{2004_neal} demonstrated reversible jump MCMC for nested epidemic models with event time data augmentation. Also using reversible jump MCMC for Bayesian model selection, \cite{2012_groendyke} demonstrated exponential family random graphs that model the contact network over which disease was transmitted, while considering the covariates in the Hagelloch data set, as implemented in \pkg{epinet}. In \cite{2012_groendyke}, infection and removal times are treated as known, and exposure/transmission times are inferred. \cite{2017_meyer} also used the outbreak data as an example application for the \code{TwinSIR} individual level model in the \pkg{surveillance} package.

\paragraph*{} Here, we apply an $\mathcal{SEIR}$ TN-ILM which considers the available risk factor information and assumes the transmission network, as well as exposure, infection, and removal times to be all unknown, and to be imputed. The purpose of this analysis is to exhibit the functionality of \pkg{Pathogen.jl} using this exceptional data set, rather than to extend our understanding of the outbreak itself. The code to recreate this analysis is contained in the supplementary material in \code{1861_Measles_Hagelloch.jl}.

\paragraph*{} The Hagelloch Measles data are included in the \proglang{R} package \pkg{surveillance}. These data are further modified for ease of use with \pkg{Pathogen.jl}, and we include them as an example data set in our own package.

\paragraph*{} Our analysis of the outbreak begins by loading our required packages, and setting a seed for replicability:
\begin{Code}
using CSV, 
      DelimitedFiles, 
      Distances, 
      Random, 
      Pathogen, 
      Plots, 
      Plots.PlotMeasures, 
      DataFrames
Random.seed!(4321);
\end{Code}

\paragraph*{}
Our first step with \pkg{Pathogen.jl} is to construct a \code{Population}. We do this by loading our risk factor data, and pre-calculating a matrix of Euclidean distances between the homes of all children. We also allow transmission rates to depend on whether children were in the same household, or the same class at school. If individuals were listed at the same location, we set their distance to \code{Inf}, such that household transmission would be excluded from the Euclidean distance-based element of the infectivity kernel. For brevity we reserve the \proglang{Julia} code for construction of this distance matrix to the supplementary material. With the risk factors and distance matrix, a \code{Population} is then constructed with:
\begin{Code}
risks = CSV.read("data/measles_hagelloch_1861_risk_factors.csv", DataFrame)
pop = Population(risks, dist)
\end{Code}

\paragraph*{}
We specify observations for model fitting using the dates of prodrome, rash, and death. The date of prodrome is assumed to occur after the true onset of infectiousness. If an individual recovered, a removal time observation of the date of rash plus four days was assumed. This implies the true removal of that individual from the infectious state must have occurred before four days after displaying a rash, which is the typical maximum extent of infectiousness \citep{2020_phac}. If an individual perished before this assumed removal time, their date of death would instead be taken as their removal time. This data processing step is again reserved for supplementary material. We can construct an \code{EventObservations{SEIR}} once we have vectors of infection and removal observations with:

\begin{Code}
obs = EventObservations{SEIR}(infected, removed)
\end{Code}

\paragraph*{}
We do not exclude any cases, as has been done in some past analyses due to an isolated case following the primary outbreak. \cite{2011_groendyke} found inclusion of this particular case to affect parameter estimates of the random network model they present. In the application of a TN-ILM to this data, we expect that this case would be attributed to an external transmission source and not adversely impact model estimates, so its removal is not required. Specification of an initial infection is also not required with the TN-ILM due to it allowing for external transmission sources.

\paragraph*{} In our TN-ILM, individual disease state transitions are described by:
\begin{align}
\epsilon^{*}(i) = & \zeta \nonumber \\
\Omega_{S}(i) =& 1.0 \nonumber \\
\Omega_{T}(j) =& 1.0 \nonumber \\
\kappa(i, j) =& \alpha d_{i, j}^{-\beta} + \tau \mathcal{I}_{class}(i, j) + \rho \mathcal{I}_{house}(i, j)\label{hagelloch_kappa}\\
\Omega_{L}(i) =& \gamma \nonumber \\
\Omega_{R}(i) =& \eta \nonumber
\end{align}

\noindent where $d_{i, j}$ is the Euclidean distance between individuals $i$ and $j$, $\mathcal{I}_{class}(i, j)$ and $\mathcal{I}_{house}(i, j)$ are indicator functions for whether $i$ and $j$ are members of the same classroom, and same household, respectively. This model structure is specified as 

\begin{Code}
function _constant(params::Vector{Float64}, pop::Population, i::Int64)
  return params[1]
end

function _one(params::Vector{Float64}, pop::Population, i::Int64)
  return 1.0
end

function _powerlaw_plus(params::Vector{Float64}, pop::Population, i::Int64, k::Int64)
  return params[1] * pop.distances[k, i][1]^(-params[2]) +
         params[3] * pop.distances[k, i][2] +
         params[4] * pop.distances[k, i][3]
end

rf = RiskFunctions{SEIR}(_constant, # sparks function
                         _one, # susceptibility function
                         _powerlaw_plus, # infectivity function
                         _one, # transmissability function
                         _constant, # latency function
                         _constant) # removal function
\end{Code}

\paragraph*{} Weakly informative priors that restrain parameters to positive values were selected for all variables for the purposes of fitting this model using MCMC, with:
\begin{align}
    \zeta \sim& \text{Uniform}(0.0, 0.1), \nonumber \\
    \alpha \sim& \text{Uniform}(0.0, 7.0), \nonumber \\
    \beta \sim& \text{Uniform}(0.0, 7.0), \nonumber \\
    \tau \sim& \text{Uniform}(0.0, 1.0), \nonumber \\
    \rho \sim& \text{Uniform}(0.0, 1.0), \nonumber \\
    \gamma \sim& \text{Uniform}(0.0, 1.0), \nonumber \\
    \eta \sim& \text{Uniform}(0.0, 1.0). \nonumber \\
\end{align}

\noindent Uniform prior distributions were specified for the delay between true infection and removal times and infection and removal observations of Uniform$(0.0, 3.0)$ and Uniform$(0.0, 2.0)$, respectively. The prior distribution for exposure times is specified relative to the latent infection time as having occurred Uniform$(5.0, 14.0)$ days earlier. A flat uniform prior was assumed for the transmission network.

Using \pkg{Pathogen.jl}, we specify these prior distributions with:
\begin{Code}
rpriors = RiskPriors{SEIR}([Uniform(0.0, 0.1)],
                           UnivariateDistribution[],
                           [Uniform(0.0, 7.0)
                            Uniform(0.0, 7.0)
                            Uniform(0.0, 1.0)
                            Uniform(0.0, 1.0)],
                           UnivariateDistribution[],
                           [Uniform(0.0, 1.0)],
                           [Uniform(0.0, 1.0)])

ee = EventExtents{SEIR}((5.0, 14.0), 3.0, 2.0)
\end{Code}

\noindent We next construct an \code{MCMC\{SEIR\}} object, and initialize it by selecting the best parameter set from 100k initialization attempts. From the selected initial parameter values, we then perform 200k MCMC iterations. For this example, event time data augmentation is completed in 10 blocks, with a transition kernel variance set to $1.0$ for each event time.

\begin{Code}
mcmc = MCMC(obs, ee, pop, rf, rpriors)
start!(mcmc, attempts=100000)
iterate!(mcmc, 200000, 1.0, condition_on_network=true, event_batches=10)
\end{Code}

\noindent A progress bar is displayed during MCMC initialization and iteration. It took approximately 15 minutes, and 3 hours 53 minutes to run those processes, respectively, on a computer with an Intel 2.7 GHz i7-3740QM processor, using \proglang{Julia} 1.4.2. Once completed, we can visually assess a trace plot of the TN-ILM parameters with

\begin{figure}
    \centering
    \includegraphics[width=0.6\textwidth]{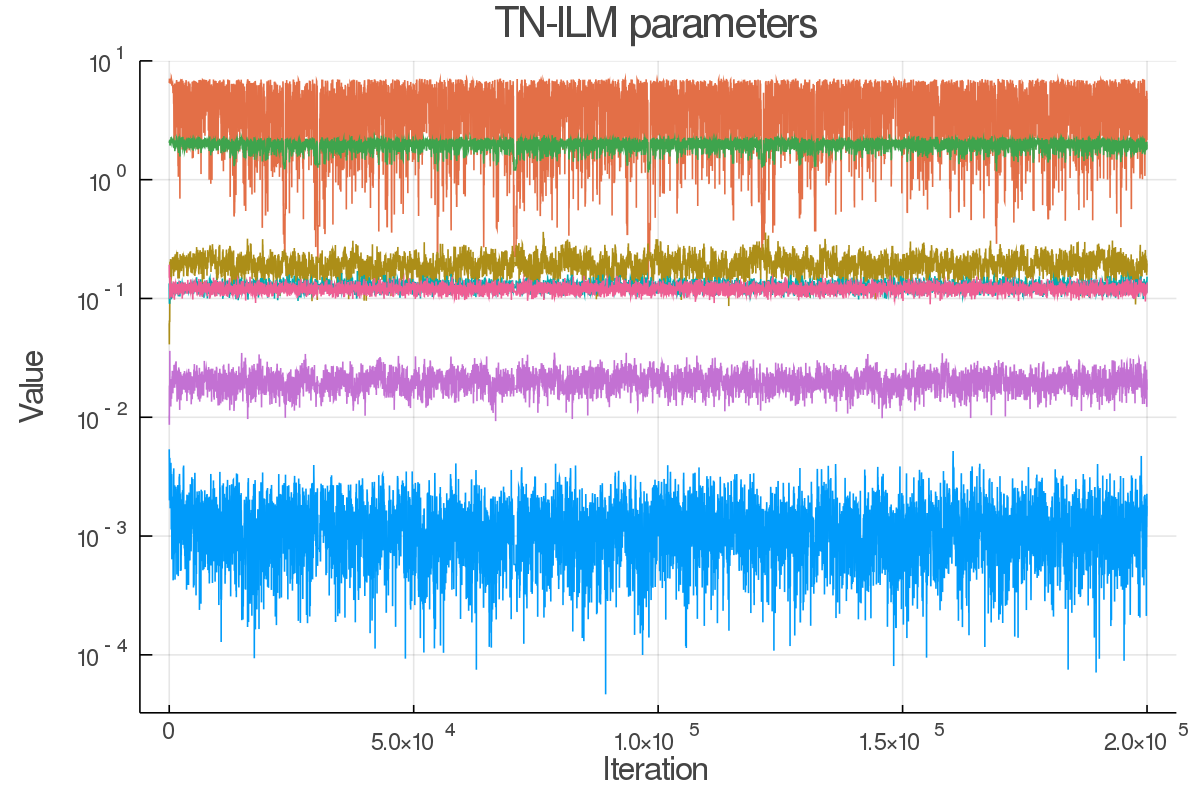}
    \caption{Trace plot after 200k iterations of a TN-ILM with 7 parameters applied to the 1861 Hagelloch measles data. Convergence was assumed to occur by 100k iterations, with a thinning rate of 50 used in selection of samples to estimate the joint posterior distribution of the TN-ILM. For display purposes, only every 20th iteration is shown in this trace plot.}
    \label{hagelloch_trace}
\end{figure}

\begin{Code}
plot(1:20:200001, mcmc.markov_chains[1].risk_parameters)
\end{Code}

\noindent which shows every 20th set of parameter values, and generates the plot seen in Figure \ref{hagelloch_trace}. The TN-ILM parameters appear to converge quickly to the posterior distribution.

\noindent Selecting a burn-in period of 100k iterations, and a thinning rate of 50 iterations, we summarize the distribution of our TN-ILM parameter with 

\begin{figure}
    \begin{center}
    \includegraphics[width=\textwidth]{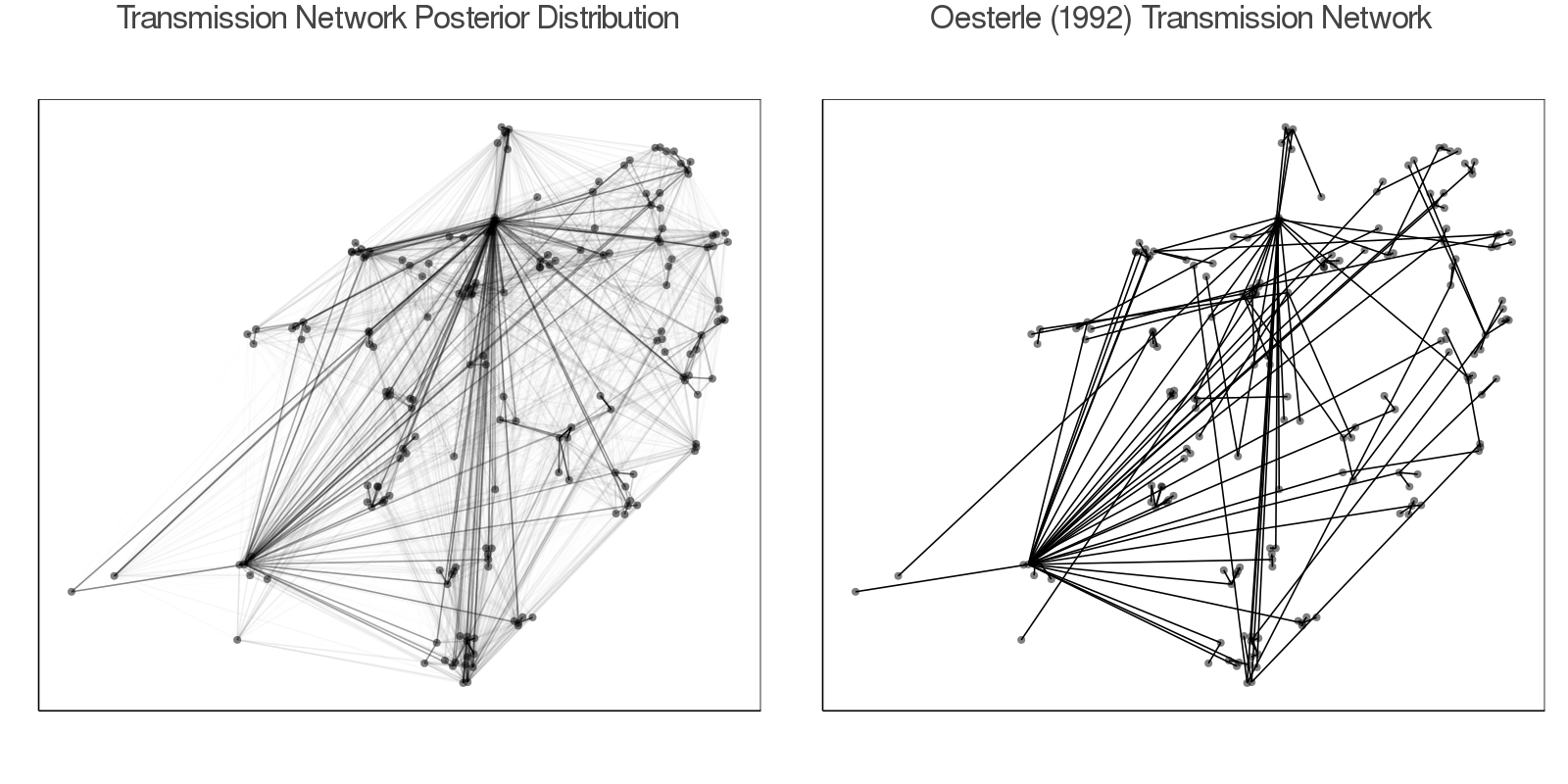}
    \includegraphics[width=\textwidth]{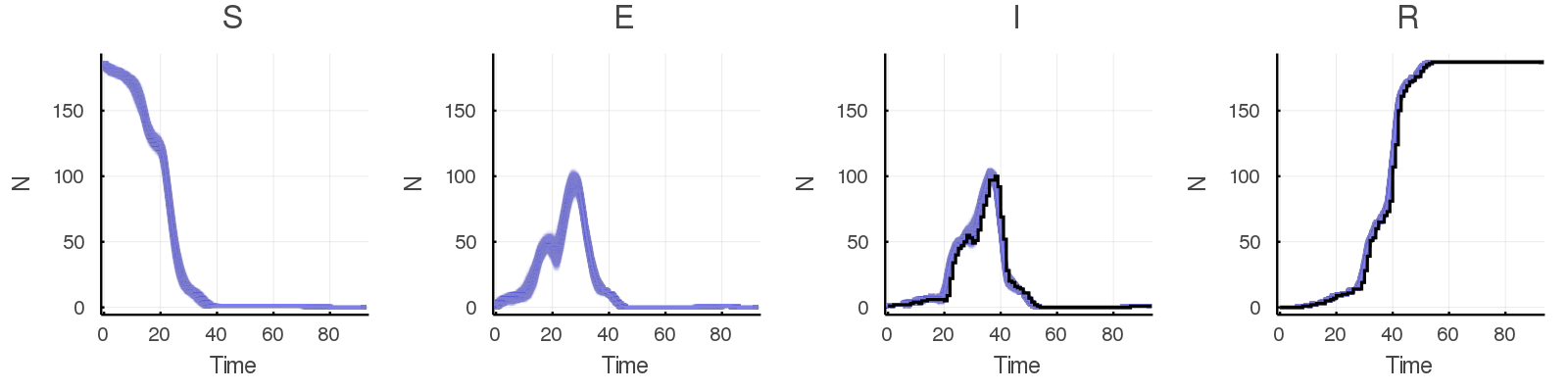}
    \end{center}
    \caption{Top: marginal posterior distribution of the transmission network is shown in contrast to the transmission network presented by \cite{1992_oesterle}. Bottom: marginal posterior distributions of the count of individuals in each disease state through time are shown, with observation of infectiousness (onset of prodrome), and observation of removal (assumed to be 4 days following onset of rash), overlaid in black.}
    \label{hagelloch_curves}
\end{figure}

\begin{Code}
summary(mcmc, burnin=100000, thin=50)
\end{Code}

which yields the posterior estimates shown in Table \ref{hagelloch_table}. We can also visually explore the posterior distribution of event times and overall epidemic curves, as well as the transmission network as we did in the simulated example in Section \ref{ex}. At the top of Figure \ref{hagelloch_curves}, we contrast the posterior distribution of the transmission network from our TN-ILM, to that suggested by \cite{1992_oesterle}. We add a small amount of variation to individual locations for display purposes only, to enable viewing of multiple individuals in the same household. The epidemic curves at the bottom of Figure \ref{hagelloch_curves} show credible bands for aggregate infection and removal times. The black lines on the infection and removal time plots indicate the observed infection and removal times. The code to recreate Figure \ref{hagelloch_curves} is included in the supplementary material.

\paragraph*{}
When comparing our transmission network posterior distribution to Oesterle's analysis, we note that Oesterle suggests individual 45 to be the source of 30 further infections, while the transmission network posterior distribution from our TN-ILM indicates an out degree of only $11.11$ for this individual. Oesterle may have been able to incorporate more contextual information into their transmission network reconstruction, resulting in this higher estimate. Our estimated out degree considers probabilities for specific transmissions, while Oesterle's transmission network reconstruction does not represent uncertainty, which also may account for some of this difference. Measles is highly infectious airborne disease, with a basic reproductive number that can vary widely based on the population, but is usually cited as in the range of 12-18 \citep{2017_guerra}. Both the value based on our transmission network reconstruction, and Oesterle's for the number of secondary transmissions for individual 45 are within a plausible range for this disease. Individual 141 is the case that has been removed in some previous analyses. Our approach yields a marginal posterior probability of this individual having an external transmission source of approximately $1.0$.

\paragraph*{}
Overall, we note that the posterior mode transmission network from our TN-ILM is in agreement with Oesterle on 103 of the 188 transmission sources. More informative prior distributions for event times, and for exposure time especially, would likely affect our transmission network reconstruction. Consideration of other risk factors, or consideration of the presently used risk factors in different forms, would also likely impact the transmission network posterior distribution.

\paragraph*{}
The epidemic curves show narrow bands for infection and removal times in the population, with higher uncertainty indicated for the exposure time. The posterior mean estimate for $\gamma$ implies a posterior mean for the latent period of $7.95$ days. The posterior mean of the incubation period is $9.42$ days. An analysis of measles data from households with two cases by \cite{2011_klinkenberg} put the mean incubation period for this disease to be slightly higher, between $11$ and $12$ days \citep{2011_klinkenberg}. The posterior mean for $\eta$ implies a mean infectious period of 8.26 days.

\paragraph*{}
The remaining parameters speak to the relative sources of infectious pressure. The infectious pressure from an infectious family member was nearly $9.73$ times higher than an infectious classmate. An infectious individual residing 15 metres away applied infectious pressure approximately equal to that of an infectious classmate. With 30 metres between residences, the infectious pressure from an infectious individual drops to approximately 25\% of that of an infectious classmate. While it was less common for an infection to have an external source, an estimate of a constant rate of $1.26 \times 10^-3$ was produced for $\zeta$. Our transmission network posterior distribution indicated an average of $7.70$ transmissions as coming from outside the observed population of 188 children in the data set.

\paragraph*{}
While modelling approaches have varied, we find these results are generally in agreement with the conclusions of other authors, being that household, then classroom transmission, were of primary importance in this outbreak.

\begin{table}[]
    \begin{center}
    \begin{tabular}{|c|c|c|c|}
    \hline
    Parameter & Mean & Variance & $95\%$ Credible Interval \\
    \hline
    $\zeta$ & $0.001261$ & $4.107\times 10^{-7}$ & $(0.0003319, 0.002840)$ \\
    $\alpha$ & $4.138$ & $3.376$ & $(0.6007, 6.872)$ \\
    $\beta$ & $1.963$ & $0.03318$ & $(1.478, 2.227)$ \\
    $\tau$ & $0.02006$ & $1.147\times 10^{-5}$ & $(0.01398, 0.02711)$ \\
    $\rho$ & $0.1952$ & $0.001172$ & $(0.1324, 0.2699)$ \\
    $\gamma$ & $0.1259$ & $8.735\times 10^{-5}$ & $(0.1084, 0.1452)$ \\
    $\eta$ & $0.1210$ & $8.102\times 10^{-5}$ & $(0.1039, 0.1391)$ \\
    \hline
    \end{tabular}
    \end{center}
    \caption{Estimates of posterior mean, posterior variance, and $95\%$ credible intervals for the 7 parameters of the TN-ILM that we applied to the Hagelloch measles data.}
    \label{hagelloch_table}
\end{table}

\section{Conclusion} \label{future}
\paragraph*{} We have introduced \pkg{Pathogen.jl}, a \proglang{Julia} package, with flexible simulation and inference capabilities for TN-ILMs. \pkg{Pathogen.jl}, currently supports $\mathcal{SEIR}$, $\mathcal{SIR}$, $\mathcal{SEI}$ and $\mathcal{SI}$ TN-ILMs. Functions of risk factors for disease state transitions within each of these model classes are effectively unrestricted, and this generality does not come at an implicit performance cost due to JIT compilation in \proglang{Julia}. With TN-ILM simulations, observational data can be generated such that realistic simulation studies can be conducted. Simulated observations such as these, or real data can be fit to TN-ILMs with \pkg{Pathogen.jl}, which performs MCMC with event time and transmission network data augmentation. For several of the types that are introduced by \pkg{Pathogen.jl}, visualization methods are also provided. These visualization methods include epidemic curves, transmission network visuals, and trace plots for MCMC.

\paragraph*{} In the future, we hope to increase the feature set of \pkg{Pathogen.jl} to include TN-ILMs incorporating time-varying disease state transition rates, and to provide an API for individual movement into and out of populations. The flexibility of \pkg{Pathogen.jl} could be explored further through the use of Artificial Neural Networks for one or more rate-describing functions, similar to what has been done with Neural Differential Equations with \pkg{DiffEqFlux.jl} \citep{2019_rackauckas}. Providing simulation and inference capabilities for phylodynamic ILMs, which incorporate densely sampled pathogen genetic sequence data, is also of particular interest. Such datasets have only relatively recently become feasible to compile with advances to sequencing technology \citep{2014_leventhal}. Pathogen genetic data can help identify likely transmission paths, and improve our understanding of key connectivity factors in disease transmission. Additional inference methods may also be added to \pkg{Pathogen.jl}, such as Approximate Bayesian Computation. Beyond these, as the \proglang{Julia} package ecosystem continues to mature, integration with other \proglang{Julia} packages will be prioritized. While not only reducing the codebase of \pkg{Pathogen.jl}, this process will present opportunity to add features, increase generality, improve performance, and improve usability of the package. Notably, there are several maturing probabilistic programming language packages in \proglang{Julia} including \pkg{Turing.jl} \citep{2018_ge}, \pkg{Soss.jl} \citep{2020_scherrer}, and \pkg{Gen.jl} \citep{2019_cusumano-towner}. The use of one of these packages could be leveraged in providing additional algorithms for fitting TN-ILMs, such as Hamiltonian Monte Carlo, variation inference, or particle filtering based approaches.

\begin{minipage}{\textwidth}
\section*{Acknowledgements}
\paragraph*{}
We thank two anonymous reviewers for their detailed and constructive feedback of this manuscript. This research was funded via a Highly Qualified Personnel HQP scholarship from the Ontario Ministry of Agriculture, Food and Rural Affairs (OMAFRA) / University of Guelph  Partnership, as well as from Dr. Feng’s and Dr. Deardon’s Natural Sciences and Engineering Research Council of Canada (NSERC) Discovery Grants.
\end{minipage}
\bibliography{ref}

\begin{thebibliography}{37}
\newcommand{\enquote}[1]{``#1''}
\providecommand{\natexlab}[1]{#1}
\providecommand{\url}[1]{\texttt{#1}}
\providecommand{\urlprefix}{URL }
\expandafter\ifx\csname urlstyle\endcsname\relax
  \providecommand{\doi}[1]{doi:\discretionary{}{}{}#1}\else
  \providecommand{\doi}{doi:\discretionary{}{}{}\begingroup
  \urlstyle{rm}\Url}\fi
\providecommand{\eprint}[2][]{\url{#2}}

\bibitem[{Abdulkareem \emph{et~al.}(2020)Abdulkareem, Augustijn, Filatova,
  Musial, and Mustafa}]{2020_abdulkareem}
Abdulkareem SA, Augustijn EW, Filatova T, Musial K, Mustafa YT (2020).
\newblock \enquote{Risk perception and behavioral change during epidemics:
  Comparing models of individual and collective learning.}
\newblock \emph{PloS One}.
\newblock \doi{10.1371/journal.pone.0226483}.

\bibitem[{Almutiry \emph{et~al.}(2020)Almutiry, Deardon, and {Warriyar K.
  V.}}]{EpiILMCT}
Almutiry W, Deardon R, {Warriyar K V} V (2020).
\newblock \emph{\pkg{EpiILMCT}: Continuous time distance-based and
  network-based individual level models for epidemics}.
\newblock \proglang{R} package version 1.1.6,
  \urlprefix\url{https://CRAN.R-project.org/package=EpiILMCT}.

\bibitem[{Besan\c{c}on \emph{et~al.}(2019)Besan\c{c}on, Anthoff, Arslan, Byrne,
  Lin, Papamarkou, and Pearson}]{2019_besancon}
Besan\c{c}on M, Anthoff D, Arslan A, Byrne S, Lin D, Papamarkou T, Pearson J
  (2019).
\newblock \enquote{\pkg{Distributions.jl}: Definition and modeling of
  probability distributions in the \code{JuliaStats} ecosystem.}
\newblock \emph{arXiv preprint arXiv::1907.08611}.

\bibitem[{Bezanson \emph{et~al.}(2018)Bezanson, Chen, Chung, Karpinski, Shah,
  Vitek, and Zoubritzky}]{2018_bezanson}
Bezanson J, Chen J, Chung B, Karpinski S, Shah VB, Vitek J, Zoubritzky L
  (2018).
\newblock \enquote{\proglang{Julia}: Dynamism and performance reconciled by
  design.}
\newblock \emph{Proceedings of the ACM on Programming Languages},
  \textbf{2}(OOPSLA), 120:1--120:23.
\newblock ISSN 2475-1421.
\newblock \doi{10.1145/3276490}.

\bibitem[{Bezanson \emph{et~al.}(2017)Bezanson, Edelman, Karpinski, and
  Shah}]{2017_bezanson}
Bezanson J, Edelman A, Karpinski S, Shah VB (2017).
\newblock \enquote{\proglang{Julia}: A fresh approach to numerical computing.}
\newblock \emph{SIAM Review}, \textbf{59}(1), 65--98.
\newblock \doi{10.1137/141000671}.

\bibitem[{Breloff(2015)}]{plots}
Breloff T (2015).
\newblock \enquote{\pkg{Plots.jl}: Powerful convenience for visualization in
  \proglang{Julia}.}
\newblock \urlprefix\url{http://docs.juliaplots.org/}.

\bibitem[{Cusumano-Towner \emph{et~al.}(2019)Cusumano-Towner, Saad, Lew, and
  Mansinghka}]{2019_cusumano-towner}
Cusumano-Towner MF, Saad FA, Lew AK, Mansinghka VK (2019).
\newblock \enquote{\pkg{Gen}: A general-purpose probabilistic programming
  system with programmable inference.}
\newblock In \emph{Proceedings of the 40th ACM SIGPLAN Conference on
  Programming Language Design and Implementation}, PLDI 2019, pp. 221--236.
  ACM, New York, NY, USA.
\newblock ISBN 978-1-4503-6712-7.
\newblock \doi{10.1145/3314221.3314642}.

\bibitem[{Deardon \emph{et~al.}(2010)Deardon, Brooks, Grenfell, Keeling,
  Tildesley, Savill, Shaw, and Woolhouse}]{2010_deardon}
Deardon R, Brooks SP, Grenfell BT, Keeling MJ, Tildesley MJ, Savill NJ, Shaw
  DJ, Woolhouse ME (2010).
\newblock \enquote{Inference for individual-level models of infectious diseases
  in large populations.}
\newblock \emph{Statistica Sinica}, \textbf{20}(1), 239--261.

\bibitem[{Ge \emph{et~al.}(2018)Ge, Xu, and Ghahramani}]{2018_ge}
Ge H, Xu K, Ghahramani Z (2018).
\newblock \enquote{\pkg{Turing}: A language for flexible probabilistic
  inference.}
\newblock In \emph{International Conference on Artificial Intelligence and
  Statistics, {AISTATS} 2018, 9-11 April 2018, Playa Blanca, Lanzarote, Canary
  Islands, Spain}, pp. 1682--1690.
\newblock \urlprefix\url{http://proceedings.mlr.press/v84/ge18b.html}.

\bibitem[{Gillespie(1977)}]{1977_gillespie}
Gillespie DT (1977).
\newblock \enquote{Exact stochastic simulation of coupled chemical reactions.}
\newblock \emph{The Journal of Physical Chemistry}, \textbf{81}(25),
  2340--2361.
\newblock \doi{10.1021/j100540a008}.

\bibitem[{Groendyke and Welch(2018)}]{2018_Groendyke&Welch}
Groendyke C, Welch D (2018).
\newblock \enquote{\pkg{epinet}: An \proglang{R} package to analyze epidemics
  spread across contact networks.}
\newblock \emph{Journal of Statistical Software, Articles}, \textbf{83}(11),
  1--22.
\newblock \doi{10.18637/jss.v083.i11}.

\bibitem[{Groendyke \emph{et~al.}(2011)Groendyke, Welch, and
  Hunter}]{2011_groendyke}
Groendyke C, Welch D, Hunter DR (2011).
\newblock \enquote{{B}ayesian inference for contact networks given epidemic
  data.}
\newblock \emph{Scandinavian Journal of Statistics}, \textbf{38}(3), 600--616.
\newblock \doi{10.1111/j.1467-9469.2010.00721.x}.

\bibitem[{Groendyke \emph{et~al.}(2012)Groendyke, Welch, and
  Hunter}]{2012_groendyke}
Groendyke C, Welch D, Hunter DR (2012).
\newblock \enquote{A network-based analysis of the 1861 {H}agelloch measles
  data.}
\newblock \emph{Biometrics}, \textbf{68}(3), 755 -- 765.
\newblock ISSN 0006341X.
\newblock \doi{10.1111/j.1541-0420.2012.01748.x}.

\bibitem[{Guerra \emph{et~al.}(2017)Guerra, Bolotin, Lim, Heffernan, Deeks, Li,
  and Crowcroft}]{2017_guerra}
Guerra FM, Bolotin S, Lim G, Heffernan J, Deeks SL, Li Y, Crowcroft NS (2017).
\newblock \enquote{The basic reproduction number ({R0}) of measles: a
  systematic review.}
\newblock \emph{The Lancet Infectious Diseases}, \textbf{17}(12), e420--e428.
\newblock \doi{10.1016/S1473-3099(17)30307-9}.

\bibitem[{Hastings(1970)}]{1970_hastings}
Hastings WK (1970).
\newblock \enquote{{{M}onte {C}arlo sampling methods using {M}arkov chains and
  their applications}.}
\newblock \emph{Biometrika}, \textbf{57}(1), 97--109.
\newblock ISSN 0006-3444.
\newblock \doi{10.1093/biomet/57.1.97}.

\bibitem[{Heinen \emph{et~al.}(1985--2020)}]{GR}
Heinen J, \emph{et~al.} (1985--2020).
\newblock \enquote{{\pkg{GR} framework}.}
\newblock \urlprefix\url{https://gr-framework.org/}.

\bibitem[{Jenness \emph{et~al.}(2018)Jenness, Goodreau, and
  Morris}]{2018_jenness}
Jenness S, Goodreau S, Morris M (2018).
\newblock \enquote{\pkg{EpiModel}: an \proglang{R} package for mathematical
  modeling of infectious disease over networks.}
\newblock \emph{Journal of Statistical Software}, \textbf{84}(8), 1--47.
\newblock ISSN 1548-7660.
\newblock \doi{10.18637/jss.v084.i08}.

\bibitem[{Klinkenberg and Nishiura(2011)}]{2011_klinkenberg}
Klinkenberg D, Nishiura H (2011).
\newblock \enquote{The correlation between infectivity and incubation period of
  measles, estimated from households with two cases.}
\newblock \emph{Journal of Theoretical Biology}, \textbf{284}(1), 52--60.
\newblock ISSN 0022-5193.
\newblock \doi{10.1016/j.jtbi.2011.06.015}.

\bibitem[{Landeros \emph{et~al.}(2018)Landeros, Stutz, Keys, Alekseyenko,
  Sinsheimer, Lange, and Sehl}]{2018_landeros}
Landeros A, Stutz T, Keys KL, Alekseyenko A, Sinsheimer JS, Lange K, Sehl ME
  (2018).
\newblock \enquote{\pkg{BioSimulator.jl}: Stochastic simulation in
  \proglang{Julia}.}
\newblock \emph{{Computer Methods and Programs in Biomedicine}}, \textbf{167},
  23--35.
\newblock \doi{10.1016/j.cmpb.2018.09.009}.

\bibitem[{Leventhal \emph{et~al.}(2014)Leventhal, G{\"u}nthard, Bonhoeffer, and
  Stadler}]{2014_leventhal}
Leventhal GE, G{\"u}nthard HF, Bonhoeffer S, Stadler T (2014).
\newblock \enquote{Using an epidemiological model for phylogenetic inference
  reveals density dependence in {HIV} transmission.}
\newblock \emph{Molecular Biology and Evolution}, \textbf{31}(1), 6--17.

\bibitem[{Meyer \emph{et~al.}(2017)Meyer, Held, and Höhle}]{2017_meyer}
Meyer S, Held L, Höhle M (2017).
\newblock \enquote{Spatio-temporal analysis of epidemic phenomena using the
  \proglang{R} Package \pkg{surveillance}.}
\newblock \emph{Journal of Statistical Software}, \textbf{77}(11), 1--55.
\newblock \doi{10.18637/jss.v077.i11}.

\bibitem[{Neal and Roberts(2004)}]{2004_neal}
Neal PJ, Roberts GO (2004).
\newblock \enquote{Statistical inference and model selection for the 1861
  {H}agelloch measles epidemic.}
\newblock \emph{Biostatistics}, \textbf{5}(2), 249--261.

\bibitem[{Oesterle(1992)}]{1992_oesterle}
Oesterle H (1992).
\newblock \enquote{Statistiche reanalyse einer masernepidemie 1861 in
  {H}agelloch.}
\newblock M.D. thesis, {Eberhard-Karls Universit\"{a}t T\"{u}bingen}.

\bibitem[{Pfeilsticker(1863)}]{1863_pfeilsticker}
Pfeilsticker A (1863).
\newblock \enquote{Beitr\"{a}ge zur pathologie der masern mit besonderer
  ber\"{u}cksichtgung der statistischen verh\"{a}ltnisse.}
\newblock M.D. thesis, {Eberhard-Karls Universit\"{a}t T\"{u}bingen}.

\bibitem[{{Public Health Agency of Canada}(2020)}]{2020_phac}
{Public Health Agency of Canada} (2020).
\newblock \enquote{Measles: symptoms and treatment.}
\newblock
  \urlprefix\url{https://www.canada.ca/en/public-health/services/diseases/measles.html}.

\bibitem[{Rackauckas \emph{et~al.}(2019)Rackauckas, Innes, Ma, Bettencourt,
  White, and Dixit}]{2019_rackauckas}
Rackauckas C, Innes M, Ma Y, Bettencourt J, White L, Dixit V (2019).
\newblock \enquote{\pkg{DiffEqFlux.jl} - a \proglang{Julia} library for neural
  differential equations.}
\newblock \emph{CoRR}, \textbf{abs/1902.02376}.
\newblock \eprint{1902.02376}, \urlprefix\url{http://arxiv.org/abs/1902.02376}.

\bibitem[{Rackauckas and Nie(2017{\natexlab{a}})}]{2017_rackauckas&nie_b}
Rackauckas C, Nie Q (2017{\natexlab{a}}).
\newblock \enquote{Adaptive methods for stochastic differential equations via
  natural embeddings and rejection sampling with memory.}
\newblock \emph{Discrete \& Continuous Dynamical Systems - B}, \textbf{22}(7),
  1--31.
\newblock \doi{10.3934/dcdsb.2017133}.

\bibitem[{Rackauckas and Nie(2017{\natexlab{b}})}]{2017_rackauckas&nie}
Rackauckas C, Nie Q (2017{\natexlab{b}}).
\newblock \enquote{\pkg{DifferentialEquations.jl} - a performant and
  feature-rich ecosystem for solving differential equations in
  \proglang{Julia}.}
\newblock \emph{Journal of Open Research Software}, \textbf{5}(1).
\newblock \doi{10.5334/jors.151}.

\bibitem[{{\proglang{R} Core Team}(2017)}]{R}
{\proglang{R} Core Team} (2017).
\newblock \emph{\proglang{R}: A language and environment for statistical
  computing}.
\newblock \proglang{R} Foundation for Statistical Computing, Vienna, Austria.
\newblock \urlprefix\url{https://www.R-project.org/}.

\bibitem[{Robert and Casella(2013)}]{2013_robert&casella}
Robert C, Casella G (2013).
\newblock \emph{{M}onte {C}arlo statistical methods}.
\newblock Springer-Verlag.
\newblock ISBN 978-1-4757-4145-2.

\bibitem[{Roberts and Rosenthal(2007)}]{2007_roberts}
Roberts GO, Rosenthal JS (2007).
\newblock \enquote{Coupling and ergodicity of adaptive {M}arkov chain {M}onte
  {C}arlo algorithms.}
\newblock \emph{Journal of Applied Probability}, pp. 458--475.
\newblock \doi{10.1239/jap/1183667414}.

\bibitem[{Ross \emph{et~al.}(2017)Ross, Baker, Parker, Ford, Mort, and
  Yates}]{2017_ross}
Ross RJ, Baker RE, Parker A, Ford M, Mort R, Yates C (2017).
\newblock \enquote{Using approximate {B}ayesian computation to quantify
  cell--cell adhesion parameters in a cell migratory process.}
\newblock \emph{NPJ systems biology and applications}, \textbf{3}(1), 1--10.
\newblock \doi{10.1038/s41540-017-0010-7}.

\bibitem[{Scherrer and Zhao(2020)}]{2020_scherrer}
Scherrer C, Zhao T (2020).
\newblock \enquote{\pkg{Soss.jl}: \code{v0.11.0}.}
\newblock \doi{10.5281/zenodo.3724489}.

\bibitem[{Shoukat \emph{et~al.}(2020)Shoukat, Wells, Langley, Singer, Galvani,
  and Moghadas}]{2020_shoukat}
Shoukat A, Wells CR, Langley JM, Singer BH, Galvani AP, Moghadas SM (2020).
\newblock \enquote{Projecting demand for critical care beds during {COVID-19}
  outbreaks in {C}anada.}
\newblock \emph{Canadian Medical Association Journal}.
\newblock \doi{10.1503/cmaj.200457}.

\bibitem[{Vahdati(2019)}]{2019_vahdati}
Vahdati A (2019).
\newblock \enquote{\pkg{Agents.jl}: Agent-based modeling framework in
  \proglang{Julia}.}
\newblock \emph{Journal of Open Source Software}, \textbf{4}(42).
\newblock \doi{10.21105/joss.01611}.

\bibitem[{{Warriyar K. V.} and Deardon(2018)}]{EpiILM}
{Warriyar K V} V, Deardon R (2018).
\newblock \emph{\pkg{EpiILM}: Spatial and network based individual level models
  for epidemics}.
\newblock \proglang{R} package version 1.4.2,
  \urlprefix\url{https://CRAN.R-project.org/package=EpiILM}.

\bibitem[{Webb(2017)}]{2017_webb}
Webb GF (2017).
\newblock \enquote{Individual based models and differential equations models of
  nosocomial epidemics in hospital intensive care units.}
\newblock \emph{Discrete \& Continuous Dynamical Systems-B}, \textbf{22}(3),
  1145--1166.
\newblock \doi{10.3934/dcdsb.2017056}.

\end{thebibliography}
\end{document}